\documentclass[a4paper,12pt]{article}

\usepackage[a4paper]{geometry}
\usepackage[utf8]{inputenc} 
\usepackage[draft]{hyperref}
\usepackage{float}
\usepackage{amssymb}
\usepackage{amsmath}
\usepackage{amsthm}
\usepackage{url}
\usepackage{amsfonts}
\usepackage{graphicx, float, tabularx}
\usepackage{color}
\usepackage{enumerate}
\usepackage[american]{babel}
\usepackage{stackrel}
\usepackage[numbers,sort,compress]{natbib}
\newcommand{\beq}{\begin{equation}}
\newcommand{\eeq}{\end{equation}}

\newcommand{\LP}{\ell_\mathrm{P}}

\newcommand{\Mpl}{M_{\rm P}}

\newcommand{\Tpl}{T_{\rm P}}

\newcommand{\be}{\begin{equation}}
\newcommand{\ee}{\end{equation}}
\newcommand{\bea}{\begin{eqnarray}}
\newcommand{\eea}{\end{eqnarray}}
\newcommand{\mpl}{M_{\rm P}}

\renewcommand{\d}{\mathrm{d}}

\renewcommand{\frac}[2]{{{\displaystyle #1}\over{\displaystyle #2}}}

\author{Athanasios G. Tzikas,$^{a,b}$\thanks{E-mail: \texttt{tzikas@fias.uni-frankfurt.de}
} \and 
Piero Nicolini,$^{a,b}$\thanks{E-mail: \texttt{nicolini@fias.uni-frankfurt.de}
} \and 
Jonas Mureika,$^{c}$\thanks{E-mail: \texttt{jmureika@lmu.edu}} 
\and 
Bernard Carr$^{d}$\thanks{E-mail: \texttt{b.j.carr@qmul.ac.uk}}
\\[1ex]
\small $^{a}$Frankfurt Institute for Advanced Studies (FIAS)\\[-0.5ex]
\small Ruth-Moufang-Str.~1, D-60438 Frankfurt am Main, Germany\\[1ex]
\small $^{b}$Institut f\"{u}r Theoretische Physik, Goethe-Universit\"{a}t Frankfurt am Main\\[-0.5ex]
\small Max-von-Laue-Str.~1, D-60438 Frankfurt am Main, Germany\\[1ex]
\small $^{c}$Department of Physics, Loyola Marymount University 
\\[-0.5ex]
\small Los Angeles, CA 90045-2659, USA\\[1ex]
\small $^{d}$School  of  Physics  and  Astronomy,  Queen  Mary  University  of  London
\\[-0.5ex]
\small Mile  End  Road,  London  E1  4NS,  UK\\[1ex]
}
\date{}
\title{Primordial black holes in a dimensionally reduced universe} 
\begin{document}
\maketitle

\vspace{0.1cm}

\begin{abstract}
\noindent  
%\PN{This is the last thing to review}
% {
%\small
%In the present paper 
We investigate the  
%probability for  
 spontaneous creation of  
%neutral [?] 
primordial black holes in a lower-dimensional expanding early universe. 
%By mean of 
We use the no-boundary proposal to construct instanton solutions for both the background
%, as well as for 
and a  black hole nucleated inside this background. The resulting  
%black hole
 creation rate 
%supports the idea of the possibility of 
could lead to a significant population of primordial black holes during the lower dimensional phase. 
%As a special result, 
We also consider the subsequent evaporation of these dimensionally reduced black holes and find 
%an inversion \textcolor{red}{[MEANING?]} of the usual black hole distribution: in contrast to of the four-dimensional picture, in the sense 
that their temperature increases with mass, whereas it decreases  with mass for 4-dimensional black holes. This means that they could leave stable sub-Planckian relics, which might in principle provide the dark matter. 
%} 

\end{abstract}

\thispagestyle{empty}
\newpage

%%% \begin{enumerate}
%%% \setlength{\itemsep}{-\parsep}
%%% %\item General relativity fails at short scales (capoziello de laurentis)
%%% %\item black holes disclose the character of QG
%%% %\item modified metrics, Planckian non geometrical approach (Casadio, Dvali, Spallucci,
%%% Hofmann and Calmet)
%%% \item Against this background gravity self completeness is the most conservative paradigm
%%%  emerging from quantum gravity considerations
%%% %\item necessity of extremal configurations as black hole remnants, suppressed back reaction
%%% \item phenomenologically produced by decay of deSitter space
%%% %\item Singularities, inner and outer solution DONE
%%% %\item Strong solutions Balasin DONE
%%% %\item BH in QG, regular and singularity avoidance DONE
%%% %\item Prior attempts of reg BH DONE
%%% %\item BH engineering DONE
%%% %\item Necessity of strong regular solutions DONE
%%% %\item Recently Zerbini DONE
%%% %\item Scheme of the paper DONE
%%% %\item solution construction
%%% %\item comment on eq of state
%%% \end{enumerate}
%%% 
%%% \newpage

%\tableofcontents

%\newpage

\section{Introduction}

Gravity is 
%probably 
the most familiar and least understood fundamental interaction. We experience its influence in everyday life but  progress towards its full understanding has been slow. On the experimental side, the accuracy of the  value of the Newton's constant $G_{\mathrm{N}}\simeq 6.67384(80)\times 10^{-11} \ \mathrm{N} (\mathrm{m}/\mathrm{kg})^2$ \cite{MNT16} has barely improved since the time of Cavendish \cite{Gil97}. On the theoretical side, the situation is no more promising. General relativity suffers from bad short-distance behavior and 
%fails to be predictive
may need to be modified at large distances unless one invokes {\it ad hoc} dark sectors.
% [OK?]. 
Black holes are plagued with paradoxes connected to their information content \cite{tGR16}. The quantization of gravity as a standard quantum field theory has been frustrated by its non-renormalizable character, as 
%it is evident from
indicated by  the dimensionality of its coupling constant, $G_{\mathrm{N}}=\mpl^{-2}$ in natural units. Alternative quantization schemes,
% most notably 
such as superstring theory, loop quantum gravity and asymptotically safe gravity, have made major breakthroughs 
%in the field but none of them has yet reached a 
but there is no consensus on which approach is correct. 

Against this background, there has been a radical proposal to circumvent the difficulties in reconciling gravity and quantum mechanics. Following the seminal paper by 't Hooft  \cite{tHo93}, one can conjecture that the Universe 
%``actually 
behaves two-dimensionally at the Planck scale, as a consequence of spontaneous dimensional reduction. From this perspective, the non-renormalizabity of gravity would be an apparent low energy feature that shows up only when the Universe ``oxidates'' to the standard, infrared $(3+1)$-dimensional manifold. In other words, general relativity would be just an effective description of a fundamental theory of quantum gravity governed by a dimensionless coupling constant $G_{(2)}$.

To understand the meaning of a universe behaving two-dimensionally, one has to invoke an alternative definition of dimension.  The study of quantum fluctuations at the Planck scale suggests that fractal
% indicators are 
dimensionality is the most appropriate concept in the early Universe.
% in order to speak of Universe dimension. For instance, 
The spectral dimension is
% one of such 
an indicator well suited for this purpose and its computation has been the focus of several investigations based on a variety of quantum spacetime models. For example, causal dynamical triangulation \cite{AJL05}, loop quantum gravity \cite{Mod09}, asymptotically safe gravity \cite{LaR05,ReS11}, noncommutative geometry \cite{MoN10}, multi-fractal geometry \cite{Cal12}, modified dispersion relations \cite{AAG13+} all 
%confirm 
suggest a continuous dimensional flow to two dimensions. 
It has also been shown that, due to their scaling dimension, unparticles -- a sector of massive particles beyond the Standard Model (SM), conjectured to play a fundamental role in
the discretization of the Universe at the Planck scale -- 
%are the form of matter that 
can naturally arise in a dimensionally-reduced, fractalized universe \cite{NiS11}; see \cite{Car17} for a recent review.

\begin{figure}[t!] 
\begin{center}
\includegraphics[width=0.9 \textwidth]{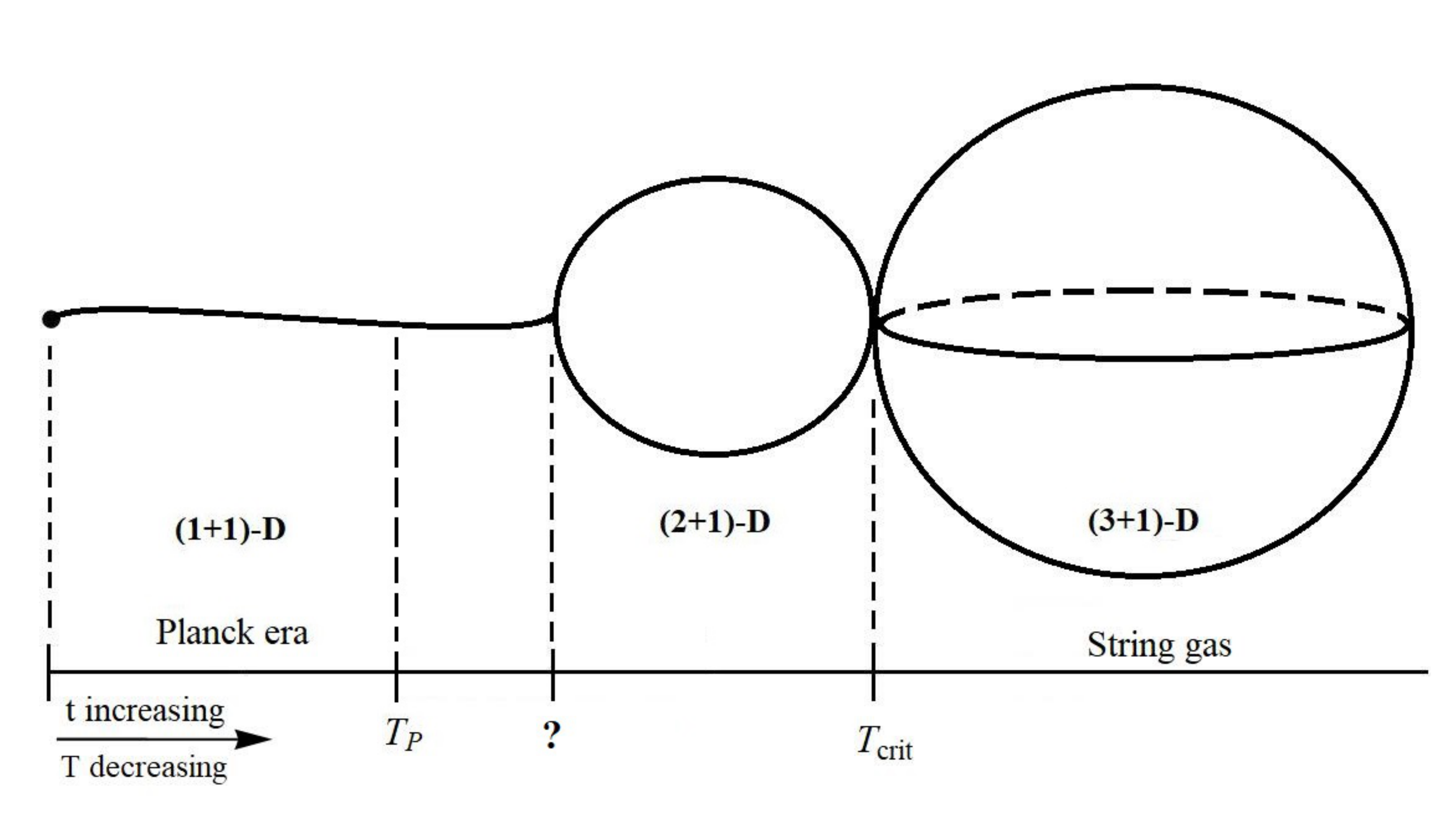}
\caption{Illustrating the Universe's dimensional oxidation. During the Planck era it behaves two-dimensionally and below some critical temperature, $T_\mathrm{crit}$, it  has the conventional four-dimensional form.
 From a string theory perspective, there may exist an intermediate three-dimensional phase with
 %phase transition at a critical temperature 
$T_\mathrm{crit}<
%T_\mathrm{Hagedorn }
 T_\mathrm{Hag} \ll \Tpl \simeq 1.42\times 10^{32}$ K  during which it
%the Universe to the conventional four dimensional Universe,
is filled with a gas of strings \cite{AtW88}.
% [INFLATION?]
%[REPHRASED]
% \textcolor{red}{[RELEVANT DISCUSSION IN SEC. 5 MUST BE  BROUGHT FORWARD.]} 
}  
\label{fig:Dimevol}
\end{center}
\end{figure}

The interpretation of the above results is as follows: while the number of spacetime dimension remains the usual topological one, \textit{i.e.} $(3+1)$, matter fields and gravity itself may not be able to perceive all of them due to the local loss of resolution expected at Planckian energies.    
%For phenomenological such a dimensionally reduced phase may have existed prior the beginning of inflation, \textit{i.e.}, for a age of the Universe smaller than $10^{-36}$ s.
These results have led to a resurgence of interest in $(1+1)$-dimensional black holes  
in relation to such quantum
gravitational characteristics as singularity avoidance \cite{MuN11}, gravitational ultraviolet self-completeness \cite{MuN13,CMN15} and the recently proposed black hole chemistry \cite{FMM15}.

In light of these developments,
%such a background work, in the present paper, we 
this paper addresses the issue of primordial black hole (PBH) production in a dimensionally reduced universe. 
%Customarily 
One usually expects gravitational collapse into PBHs to have occurred as a consequence of  large density fluctuations in the early Universe \cite{CaH74}. After their evaporation some part of the initial PBH mass might survive as
%have turned into 
a cold remnant, thereby contributing to 
%becoming a possible 
the dark matter  \cite{macgibbon}.
% component. It has, however, been noted that 
However, in the lower-dimensional
case, such a scenario needs to be revised \cite{Mur12}. First, in $(2+1)$-dimensions, black hole solutions of the Einstein equations exist only
in Anti-de Sitter space, 
%for negative values of the cosmological constant, 
suggesting that the aforementioned 
%Universe
 oxidation from two to four dimensions might have been a non-analytic phase transition, as indicated in Fig. \ref{fig:Dimevol}. 
%%% For consistency, it %such a transition 
%%% must have occurred no later than the beginning of inflation, \textit{i.e.}, in a time window  $10^{-43}\ \mathrm{s}<t<10^{-36}\ \mathrm{s}$, as indicated in Fig. \ref{fig:Dimevol}. {\textcolor{red}{[BUT IT MUST EXTEND UNTIL AFTER INFLATION]}
 Second, $(1+1)$-dimensional dilaton gravity black holes radiate with power proportional to their mass squared.
Therefore, in marked contrast with their $(3+1)$-dimensional counterparts, lighter black holes would be more stable than heavier ones. This idea has been expanded upon in the context of the Bose-Einstein condensate corpuscular picture, wherein the end stage of black hole evaporation obeys a dimensionally-reduced behavior \cite{rcjm1}.  The probability of lower-dimensional
black hole formation was also estimated using the horizon wavefunction approach
\cite{rcjm2}, a method which was recently applied to production of PBHs in $(3+1)$-dimensions
\cite{casadiopbh}.
% in two dimensions. 

 In order to estimate the PBH nuclear rate, we follow the gravitational instanton approach proposed by Mann \& Ross \cite{MaR95} and Bousso \& Hawking \cite{BoH96}. 
 %In the Euclidean quantum gravity formulation, it is possible to perform a semiclassical analysis by a suitable estimate of the gravitational path integrals \cite{GiH93}. 
% It is well known that a pure theory of quantum gravity has not yet been developed, due to its non-renormalisable character in (3+1)-dimensions. Nevertheless, attempts to approach its quantisation  have been made and one popular among them is called Euclidean quantum gravity \cite{GiH93}. This is a semi-classical \footnote{In the semi-classical approach we treat spacetime geometry classically, while the various propagating fields quantum mechanically.} theory formulated by gravitational path integrals and helps us investigate non-pertubative effects, such as spontaneous black hole formation. 
%\par 
Within the Euclidean quantum gravity formulation \cite{GiH93}, the Hartle-Hawking no-boundary proposal \cite{HaH83} implies that a spacetime can be represented quantum-mechanically by 
%means of 
a  wavefunction  $\Psi \,$, 
%which is 
defined by a path integral over positive-definite Euclidean metric configurations.
%They use
In a semiclassical approach, 
%in which 
the dominant contribution comes from instantons at the saddle points of the Euclidean gravitational action 
%$ S_{\mathrm{E}}$. In this case, and  the wavefunction becomes
and $\Psi$ is determined by the 
% \approx e^{-\mathcal{I}}$ where $\mathcal{I} \,$ is 
the instanton action  \cite{GiH79}. 
Squaring this then gives 
%a probability, of a specific universe. Thus so 
the relative probability ($\Gamma$)
%one can define the black hole creation rate, $\Gamma \,$, as the probability ratio 
of two universes \cite{
%BoH96,
MaN11}, one representing an empty background
 %$(\mathcal{I}_{\mathrm{bg}})$ 
and the other a black hole inside such a background.  
%This indicates how suppressed ($\Gamma < 1$) or favored ($\Gamma >1$)  black hole fomation is in such a universe. 
%From the above formalism, In the latter case, 
For $\Gamma >1$, 
%one can infer that 
de Sitter space is  quantum mechanically unstable, despite being classically stable.
\if
For example, 
%one has 
 the Schwarzschild-de Sitter-Nariai in $(3+1)$-dimensions has
%creation rate this becomes
$\Gamma = e^{-\pi/\Lambda_\mathrm{c}}$
in geometrical units with $G_{\mathrm{N}}=1$  \cite{BoH95}. 
\fi
 $\Gamma $ depends exponentially on the inverse of the cosmological constant \cite{BoH95,DNT18}, so PBH production  
%Therefore the  rate \eqref{a4}  must be 
is strongly suppressed for the presently observed value of this constant ($\Lambda_\mathrm{c}  \sim 10^{-120})$ but 
%becomes 
non-negligible for $\Lambda_\mathrm{c}  \sim 1 \,$.
%so Planckian PBHs might have been produced prolifically from  the decay of de Sitter space.}  
%The radius of these PBHs is given by $r=\frac{1}{\sqrt{\Lambda}} \,$. Hence,  Planckian PBHs have a good probability to form spontaneously inside an inflationary and (3+1)-dimensional universe.  Our goal is to repeat this process for a (1+1)-dimensional universe. 
%\par   In recent years, many theoretical evidence \cite{AtW88,tHo93,AJL05,Mod09,Car09,MoN10,NiS11,MuS11,Mur12,Rin12,MuN13,Car16} illustrate the possibility that our world may not behave as a (3+1)-dimensional spacetime when we are working at high energy scales or at short distances and, instead of introducing extra dimensions, physical systems lower their dimensionality at 2. That means, there might existed a pre-inflationary phase of the  early universe, where the universe itself started as an effective (1+1)-dimensional spacetime, then became a (2+1)-dimensional surface and finally the (3+1)-dimensional universe that we live today. The word "effective" means that we can describe this (sub-)Planckian phase in the framework of a lower dimensional gravity \cite{CMN15}, instead of ordinary or higher dimensions. 

%On this basis, 
  
 In this paper, we use
%we aim to exploit
 the instanton formalism to  calculate the PBH production rate in a dimensionally-reduced universe. However, this raises several important issues, none of which is fully resolved at this stage. One needs to distinguish these issues, since they involve different  conceptual problems. 
The first issue concerns  the nature of the transition from the $1+1$ to $3+1$ phase. 
%[ADDED] Another important issue is whether 
One possibility is that there was an intermediate $2+1$ phase. For example,
Atick and Witten have shown that a gas of strings, heated up to a critical temperature, $T_\mathrm{crit}$, below the Hagedorn temperature, $T_\mathrm{Hag}$, undergoes a phase transition to an effective $2+1$ phase \cite{AtW88}. This is illustrated in Fig.~\ref{fig:Dimevol} but does not address the question of how to describe the black hole during this intermediate stage. The important point is that this scenario allows the dimensional transition to occur  below the Planck temperature.

 The second issue concerns the behavior of the  cosmological  constant during the transition. One must distinguish between the 2D cosmological constant $\Lambda$ and the current 4D cosmological constant  $\Lambda_c$. In this context, one  might expect the $1+1$ phase to be inflationary, 
%\textcolor{red}{[CHECK]}  
since one can always
%presumably 
choose a spacelike slicing in which the metric has an exponentially expanding form  \cite{ChM95}.
%,  as with the $3+1$ de Sitter solution. The issue of whether $1+1$ models can expand exponentially 
\if
This issue has been studied by Chan and Mann \cite{ChM95}, who argue that a pure-radiation model is static and that one needs some exotic matter content to drive inflation.
% [EG. NEGATIVE $\Lambda$ BUT BETTER TO DISCUSS THIS LATER].
%However, there is no such content in our model. ]
%\PN{yes}
\fi
However, in this case, is the  $1+1$ phase complete before the conventional $3+1$ inflationary phase or does it {\it replace} it? 
One might be reluctant to replace it because then one could lose some of  the attractions of the standard scenario (eg. the form of the density fluctuations). On the other hand, if the $1+1$ phase ends before $3+1$ inflation, there would be no observational consequences of $1+1$ PBH production. 

The third issue concerns the black hole solution itself. One must understand how a $1+1$ black hole turns into a $3+1$ one
and this is non-trivial. Possibly the region close to the  black hole remains ($1+1$)-dimensional after the cosmological transition. Indeed, this is implicit in our discussion of the evaporation of a $1+1$ black hole in a $3+1$ cosmological background. This would entail the introduction of a spatial inhomogeneity but that is implicit in PBH production anyway.

%[THIS MUST BE IMPROVED BUT WE NEED TO SAY SOMETHING ABOUT THIS.]  

The paper is organized as follows. In Sec. \ref{sec:dimred} we review dilaton/Liouville gravity as a consistent candidate to replace Einstein gravity in $(1+1)$-dimensions and we calculate the instanton for the de Sitter background. In Sec. \ref{sds_instanton} we derive two instantons for the $(1+1)$-dimensional Schwarzschild-de Sitter universe, one for a \textit{lukewarm} black hole with  a generic horizon structure
 %(a \textit{lukewarm} black hole) 
and the other  for a \textit{Nariai} black hole with a degenerate horizon. 
%(a \textit{Nariai} black hole).
 In Sec. \ref{2d_rate} we estimate the black hole production rate inside such a 
%very early, 
two-dimensional universe and in Sec. \ref{observations} we consider observational consequences of this. In Sec. \ref{conclusion} we draw some  conclusions.

\section{Dimensionally reduced de Sitter instanton}
\label{sec:dimred}

%In order to estimate the PBH nuclear rate, we follow the 
 First, we recall the gravitional instanton approach of Mann \& Ross \cite{MaR95} and Bousso \& Hawking \cite{BoH96}. 
 %In the Euclidean quantum gravity formulation, it is possible to perform a semiclassical analysis by a suitable estimate of the gravitational path integrals \cite{GiH93}. 
% It is well known that a pure theory of quantum gravity has not yet been developed, due to its non-renormalisable character in (3+1)-dimensions. Nevertheless, attempts to approach its quantisation  have been made and one popular among them is called Euclidean quantum gravity \cite{GiH93}. This is a semi-classical \footnote{In the semi-classical approach we treat spacetime geometry classically, while the various propagating fields quantum mechanically.} theory formulated by gravitational path integrals and helps us investigate non-pertubative effects, such as spontaneous black hole formation. 
%\par 
Within the Euclidean quantum gravity formulation \cite{GiH93}, the Hartle-Hawking no-boundary proposal \cite{HaH83} states that a spacetime can be treated quantum-mechanically by means of a  wavefunction  $\Psi \,$, which is defined by a path integral over positive-definite Euclidean metric configurations  $g_{\mu\nu}$:
\begin{equation} \label{a1}
\Psi= \int \mathcal{D}[g_{\mu\nu}] \ e^{-S_{\mathrm{E}}[g_{\mu\nu}]},
\end{equation} 

\noindent where $ S_{\mathrm{E}} $ is the Euclidean version of the gravitational action.  This integral does not always  converge because $ S_{\mathrm{E}} $ 
%is not always 
need not be positive. For this reason, one  uses a semiclassical approach, according to which the dominant contribution comes from instantons at the saddle points of $ S_{\mathrm{E}}$  \cite{GiH79}. In this case, apart from a prefactor, the wavefunction becomes
\begin{equation} \label{a2}
\Psi \approx e^{-\mathcal{I}},
\end{equation}  
\noindent where $\mathcal{I} \,$ is the instanton action. Squaring 
%the above wavefunction, one gets
this gives a probability,
 %of a specific universe. Thus 
 so the relative probability
%one can define the black hole creation rate, $\Gamma \,$, as the probability ratio 
of two universes \cite{
%BoH96,
MaN11}, one representing an empty background $(\mathcal{I}_{\mathrm{bg}})$ and the other a black hole inside such a background   $(\mathcal{I}_{\mathrm{bh}})$, is given by
\begin{equation}  \label{a3}
\Gamma=\exp\left[ -2(\mathcal{I}_{\mathrm{bh}}-\mathcal{I}_{\mathrm{bg}}) \right]. 
\end{equation} 
\noindent 
%Such a probability rate gives 
This indicates how suppressed ($\Gamma < 1$) or favored ($\Gamma >1$) 
 %a universe containing a black hole with respect to an empty universe
 black hole formation is in such a universe. 
%From the above formalism, 
In the latter case, one can infer that de Sitter space is  quantum mechanically unstable.
 Although $\Gamma$ is sometimes described as a ``rate'', we note that it is just a dimensionless probability.
For instance, in $(3+1)$-dimensions the Schwarzschild-de Sitter-Nariai creation rate becomes \cite{BoH95}
%For a deSitter space in  R. Bousso \& S. Hawking  calculated the creation rate of neutral primordial black holes (PBHs) in (3+1)-D, by using de Sitter spacetime as a background and  spacetime as the background containing a black hole. In Planck units, the  rate of Nariai PBHs they found reads
\begin{equation} \label{a4}
\Gamma = e^{-\pi/\Lambda_\mathrm{c}}
\end{equation}
in geometrical units with $G_{\mathrm{N}}=1$. 
\noindent  
%For consistency with observation, 
%\textcolor{red}{
The  rate
% \eqref{a4}  must be 
is strongly suppressed for the presently observed value of the cosmological constant ($\Lambda_\mathrm{c}  \sim 10^{-120})$ but becomes non-negligible for $\Lambda_\mathrm{c}  \sim 1 \,$.
This means that Planckian PBHs might have been produced prolifically from  the decay of de Sitter space.  
%The radius of these PBHs is given by $r=\frac{1}{\sqrt{\Lambda}} \,$. Hence,  Planckian PBHs have a good probability to form spontaneously inside an inflationary and (3+1)-dimensional universe.  Our goal is to repeat this process for a (1+1)-dimensional universe. 
%\par   In recent years, many theoretical evidence \cite{AtW88,tHo93,AJL05,Mod09,Car09,MoN10,NiS11,MuS11,Mur12,Rin12,MuN13,Car16} illustrate the possibility that our world may not behave as a (3+1)-dimensional spacetime when we are working at high energy scales or at short distances and, instead of introducing extra dimensions, physical systems lower their dimensionality at 2. That means, there might existed a pre-inflationary phase of the  early universe, where the universe itself started as an effective (1+1)-dimensional spacetime, then became a (2+1)-dimensional surface and finally the (3+1)-dimensional universe that we live today. The word "effective" means that we can describe this (sub-)Planckian phase in the framework of a lower dimensional gravity \cite{CMN15}, instead of ordinary or higher dimensions. 

 We now extend this approach to the lower dimensional case. Einstein  gravity in $(1+1)$-dimensions is trivial  because the Einstein tensor vanishes for every metric \cite{Bro88}.  For this reason, dilaton gravity has been the focus of a vast research activity -- see for instance \cite{Col77,Man92,MaS92,Man94,ChM95,Man95,GKV02,GrM06,MaM11} -- and offers a valuable alternative to describe systems in a dimensionally reduced spacetime. The dilaton is a scalar field that describes gravitational degrees of freedom alongside  the graviton. It is present in higher dimensional gravitational theories, most notably in Kaluza--Klein and string theory. More importantly, it has been shown that a dilaton action can be derived by taking the continuous limit, $D \rightarrow 2 \,$, of the $D$-dimensional Einstein-Hilbert action.  This guarantees a high degree of correspondence between the original fully dimensional Universe and its reduced counterpart. Specifically, the corresponding dilaton action becomes  \cite{MaR93}
\begin{equation} \label{a5}
\mathcal{S}[g_{\mu\nu}, \psi]=   \int \d^2x \sqrt{-g}\ \left[ \frac{1}{16\pi G_{(2)}}\left(  \psi R + \frac{1}{2}  (\nabla \psi)^2 - 2\Lambda \right)+ \mathcal{L}_{\mathrm{m}} \right] \, ,
\end{equation} 
\noindent  where $R \,$ is the Ricci scalar, $\psi$ is the dilaton, $\Lambda$ is the cosmological constant, $\mathcal{L}_{\mathrm{m}}$ is the matter Lagrangian and $G_{(2)}$ is  Newton's constant in $(1+1)$-dimensions. In natural units with $\hbar=c=k_{\mathrm{B}}=1$, the constant $G_{(2)}$ is dimensionless and can be set to $1$ by  convention -- see \cite{MuN13} for reference. 
%%% \textbf{[PN: can we set it to 1? If we use the convention in \cite{MuN13}, it is enough to choose $K_1=1/(2\pi)$ to have $G_{(2)}=1$. I am not setting $G_{(2)}=1$ for now.   Thanos please check where you used the value $2\pi$]}. 

The above action depends on both  the metric and the dilaton field, so
%. This means that
 its variation generates two equations of motion. The variation in $\delta\psi $ gives
\begin{equation} \label{a6}
\nabla^2 \psi =R \, ,
\end{equation} 

\noindent while the variation in $\delta g^{\mu\nu}$ gives
\begin{equation} \label{a7}
  \frac{1}{2} \nabla_{\mu} \psi \nabla_{\nu} \psi   -\frac{1}{4} g_{\mu\nu} (\nabla \psi)^2 + g_{\mu\nu} \nabla^2 \psi-\nabla_{\mu} \nabla_{\nu} \psi + \Lambda g_{\mu\nu}=8\pi G_{(2)} T_{\mu\nu},
\end{equation}

\noindent where 
%$T_{\mu\nu} \,$ is 
the stress-energy tensor  is defined as
\begin{equation} \label{a8}
T_{\mu\nu}= \frac{-2}{\sqrt{-g}} \frac{\delta(\sqrt{-g} \mathcal{L}_{\mathrm{m}})}{\delta g^{\mu\nu}} \, .
\end{equation} 

\noindent By combining  \eqref{a6} with the trace of \eqref{a7} one gets
\begin{equation} \label{a9}
R + 2\Lambda= 8\pi G_{(2)} T \, ,
\end{equation}
\noindent where $T=T^{\mu}{}_{\mu}$. 
%Equation \eqref{a9} 
This is called the Liouville field equation and governs gravity in $(1+1)$-dimensions. 
%One customarily says that  
Equation \eqref{a9} is usually regarded as the best analogue of the Einstein field equations. 
%, since it has been derived from the dropping limit of the D-dimensional Einstein-Hilbert action.  
%\par   Now that we have given  the necessary   introduction, in Sec. \ref{dS_instanton}  we construct one instanton for a (1+1)-D de Sitter universe, which represents a compact, empty and expanding background. In Sec. \ref{sds_instanton} we build two instantons for a (1+1)-D Schwarzschild-de Sitter universe, \textit{i.e.} one for a black hole having different event horizon structure from the cosmological radius (\textit{lukewarm}-like black hole) and one for a black hole having a degenerate horizon (\textit{Nariai} black hole). In Sec. \ref{2d_rate} we  calculate the creation rate of  neutral PBHs  inside this very early and 2-dimensional universe. The estimated rate states that a lower dimensional phase should enhance the population of PBHs. In Sec. \ref{conclusion} we summarise the results and draw the conclusions. We are working in natural units $(\hbar=c=k_{\mathrm{B}}=1) \,$.

%\section{De Sitter instanton}
%\label{dS_instanton}

In order to construct an instanton, we must first solve the Liouville field equation for a specific spacetime.
In analogy with the known static and stationary solutions of the $(3+1)$-dimensional case, we make the following ansatz for the line element:
\begin{equation} \label{b1}
\d s^2=- V(x)\d t^2+ \frac{\d x^2}{V(x)} \, .
\end{equation}

\noindent For vacuum solutions
% in vacuum 
$(T=0)$, Eq. \eqref{a9} takes the simple form
\begin{equation} \label{b2}
-\frac{\d^2V(x)}{\d x^2} + 2\Lambda=0 \, ,
\end{equation}

\noindent with solution 
\begin{equation} \label{b3}
V(x)=\Lambda x^2+D x + C \, ,
\end{equation} 

\noindent where $C$ and $D$ are integration constants. 
%One can see that 
Because of its dimensionality, the constant $D$ may be identified with a mass parameter in $(1+1)$-dimensions \cite{MST90}. Thus we can set it equal to zero $(D=0)$
%, because we aim to derive 
for an empty background. The value of $C \,$ is arbitrary but its sign plays an important role for the existence of horizons. 
For this spacetime there are no physical singularities, while conical singularities may be found by taking $V(x)=0\,$.
% and they signal the existence of horizons. 

%Accordingly one finds that 
The de Sitter space has a cosmological horizon  at 
\begin{equation} \label{b4}
|x_{\mathrm{c}}|=\sqrt{-\frac{C}{\Lambda}} \, ,
\end{equation}

\noindent 
%From \eqref{b4} one finds that that  
so $C$ and $\Lambda$ must have opposite signs  for real horizons to exist. As can be checked from the metric potential \eqref{b3}, a $(1+1)$-dimensional  de Sitter universe is  characterized by a negative cosmological constant $(\Lambda<0) $ \cite{FMM15}, so we can replace  $\Lambda$ with $-|\Lambda|$.
%% {\textcolor{red}{ [de Sitter in 3D has $\Lambda  > 0$, so why the difference?]}
%% {\textcolor{blue}{ [In 1+1 $\Lambda < 0$ yields deSitter topology as can be seen from the metric potential \eqref{b3}. You can also check Mann's paper \cite{FMM15}.]}
%One may see that 
Only by choosing $C>0 \,$ and $\Lambda<0 \,$  can one get a compact and expanding $(1+1)$-dimensional universe, the shape of whose potential $V$ is in analogy with the $(3+1)$-dimensional de Sitter case.  Since the value of $C$ is arbitrary and positive, we can set it to unity
%, $C=1\,$, 
without loss of generality. Thus expression \eqref{b4} for the cosmological horizon becomes  $|x_\mathrm{c}|= 1/\sqrt{|\Lambda|} \,$ and one finds 
%for the deSitter 
\begin{equation} \label{b5}
V_{\mathrm{dS}}(x)=1-|\Lambda| x^2 \, .
\end{equation} 

\begin{figure}[h!] 
\begin{center}
\includegraphics[width=0.7 \textwidth]{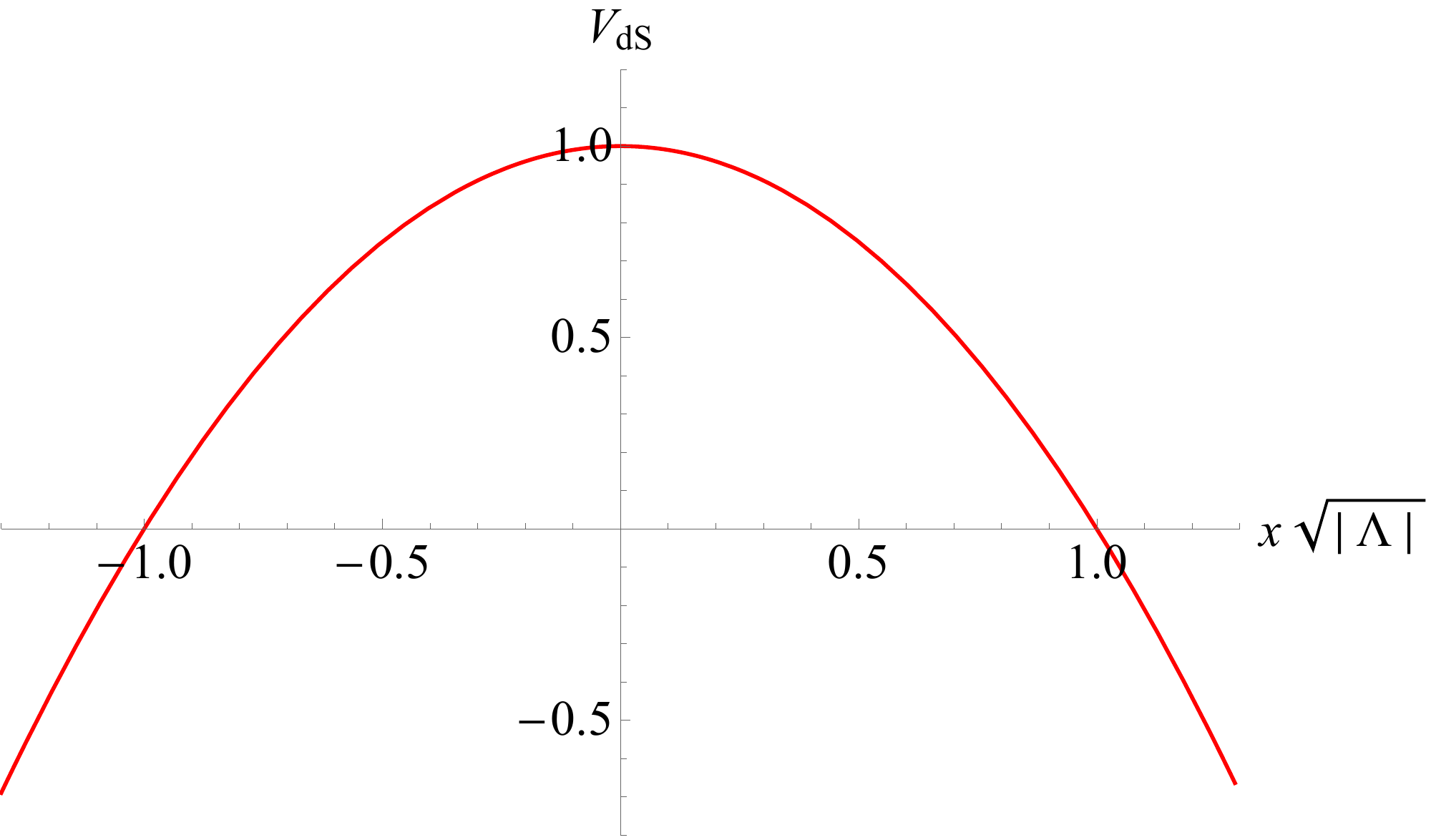}
\caption{The function $V_{\mathrm{dS}}(x) \,$ \textit{vs} $x\sqrt{|\Lambda|}$. }
\label{fig:Fig. 1}
\end{center}
\end{figure}

\noindent  The  shape of \eqref{b5}  is shown in Fig. \ref{fig:Fig. 1}. 
%% in arbitrary length units $(\ell)$ \textbf{[PN: Thanos, I think we do not need to introduce an extra scale $\ell$. Can you simply plot $V$ vs $x$ for $\Lambda=1$, i.e., $V$ vs $x\sqrt{\Lambda}$ ]}.
 For the regularity of the instanton, we analytically continue the time, $\tau = it$.  Hawking radiation is associated with the periodicity of $\tau$,  the black hole temperature being the inverse of the period. This is  $\beta = 2 \pi/ \kappa$, where $\kappa$ is the surface gravity, given by 
\begin{equation}
\kappa=\frac{1}{2} \left| \frac{\d V}{\d x} \right|_{x=x_\mathrm{c}} = |\Lambda  x_\mathrm{c}| = \sqrt{| \Lambda |} \, ,
\end{equation}
using eqs.~(\ref{b4}) and (\ref{b5}). 
%The periodicity of $\tau$  gives the Hawking temperature in terms of the inverse of the period 
%$\beta$, namely $\beta=2\pi/ \kappa \,$.
%, where $\kappa \,$ is the surface gravity  $\kappa=\frac{1}{2} \left| \frac{dV}{\d x} \right|_{x=x_\mathrm{c}} \,$.
 So in our case, the period is 
\begin{equation} \label{b6}
\beta= \frac{2\pi}{\sqrt{|\Lambda|}} \, 
\end{equation} 
\noindent and this removes the conical singularity.  
%[REPHRASED ABOVE]
The regularity of the instanton  can be seen through an appropriate coordinate  transformation via  
%periodic 
variables $\xi$ and $\chi$, with periods $2\pi\,$ and $\pi\,$, respectively: 
\begin{eqnarray}
\tau &=& \frac{1}{\sqrt{|\Lambda|}} \ \xi , \label{b7}\\
x &=& \frac{1}{\sqrt{|\Lambda|}} \ \cos\chi  \, .  \label{b8}
\end{eqnarray}
 Under the above transformation,  the de Sitter instanton can be written as
\begin{equation} \label{b9}
\d s^2= \frac{1}{|\Lambda|} \left( \d\chi^2+ \sin^2\chi \  \d\xi^2 \right) \, .
\end{equation} 

\noindent It can be seen that $\chi= 0$ $(x=x_{\mathrm{c}})$ is the axis of the polar coordinates $(\chi,\xi)$ and the manifold is perfectly regular there. 
%{\textcolor{red}{ [Can we transform this to the expanding 1D form with $a \propto \exp(\sqrt{|\Lambda|}t)$, as in 3D case? This is crucial for understanding PBH formation.]} \PN{We discuss this point later.} 
%\textbf{Interestingly 
As shown in Appendix \ref{Sec:appendixA}, the above line element can be cast into the two-dimensional Robertson-Walker form,
\begin{equation} \label{eq:b9}
\d s^2= - \mathrm{d} t^2 + a^2(t) \mathrm{d} x^2 \,, \quad a(t)=e^{\sqrt{|\Lambda|} t} \, ,
\end{equation} 
where the exponential scale factor corresponds to a $1+1$  inflationary model.
%$a(t)=e^{\sqrt{|\Lambda|} t} $. For the full derivation see Appendix \ref{Sec:appendixA}.}

\par The Euclidean version of the dilaton action  \eqref{a5} for this compact spacetime
 %such as above  
is 
%given by
\begin{equation} \label{b10}
\mathcal{I}=  - \int \d^2x \sqrt{g}\ \left[ \frac{1}{16\pi G_{(2)}}\left(  \psi R + \frac{1}{2}  (\nabla \psi)^2 - 2\Lambda \right)+ \mathcal{L}_m \right] \, ,
\end{equation}

\noindent so, after integrating by parts and using \eqref{a6}, 
%the calculation of 
the  $(1+1)$-dimensional de Sitter instanton-action  is
\begin{equation} \label{b11}
\mathcal{I}_{\mathrm{dS}} =   -\frac{|\Lambda|}{8\pi G_{(2)}}\int \d^2x \sqrt{g} \left( \frac{1}{2} \psi +1\right). 
\end{equation} 
%% \noindent {\textcolor{red}{ [No $\nabla \Psi $ term?]} {\textcolor{blue}{ [No $\nabla \Psi $-term exists because we integrated by parts the $(\nabla \Psi)^2 $-term in the initial action and we used eq.(6) and eq.(9) in vacuum, i.e, $\nabla^2 \Psi = R = 2|\Lambda| $.  ]} 
By solving the equations of motion \eqref{a6}, the dilaton becomes
%\begin{equation} \label{b12}
%\psi =\psi _0 ^{(\mathrm{dS})}+ \tanh^{-1}(x \sqrt{|\Lambda|}) - \ln  \left( 1 - |\Lambda| x^2 \right)  \, ,
%\end{equation} 
%% \if
%% where $\psi _0 ^{(\mathrm{dS})}$ is an integration constant.
%% \textbf{[PN: I do not like this notation that comes from the computer science community. Thanos, please check that what I am writing below in place of the above equation is correct]
%% \fi
%% \begin{equation} \label{b12}
%% \psi =\psi _0 ^{(\mathrm{dS})} + \mathrm{arctanh}(x \sqrt{|\Lambda|}) - \ln  \left( 1 - | 
%% \Lambda| x^2 \right)  \,  
%% \end{equation} 
%% 
%%  {\textcolor{blue}{ [There is a $(-)$ typo mistake in eq.(21). The correct form is:
\begin{equation} 
\label{b12}
\psi =\psi _0 ^{(\mathrm{dS})} - \mathrm{arctanh}(x \sqrt{|\Lambda|}) - \ln  \left( 1 - |\Lambda| x^2 \right)
\end{equation}
%%  ]}
\noindent  %We consider the dilaton as  part of the geometry and we assume that it is only a function of the position, {\it i.e.} $\psi=\psi(x) \,$. 
and by inserting 
%\eqref{b12}
this  into the instanton action \eqref{b11} we finally obtain
\begin{equation} \label{b13}
\mathcal{I}_{\mathrm{dS}}  =- \frac{\psi _0 ^{(\mathrm{dS})} + 4 - \ln4}{4 G_{(2)}} \, .
\end{equation}

\noindent {Interestingly, this result does not depend on the cosmological constant or any
%nor on 
specific energy scale,  $G_{(2)}$ being dimensionless. This means that, in contrast to the $(3+1)$-dimensional case,
 %in \eqref{a4}, 
the de Sitter instability occurs irrespective of the value of the cosmological constant. Indeed, the production rate depends only on the ratio of the mass of the tunneled object and the cosmological constant,
 %and not on the cosmological constant alone. 
so the de Sitter background will not affect the production rate.
 %As a result, 
We therefore normalize the background contribution to unity by
% a suitable choice of 
choosing the integration constant $\psi _0 ^{(\mathrm{dS})}$ so that $\mathcal{I}_{\mathrm{dS}}=0$.}
% {\textcolor{red}{[Why not $\psi _0 ^{(\mathrm{dS})} = -4$ since this agrees wih  eqn (32) when $M=0$ and $\Gamma$ depends on difference.]}
%\PN{Here it is the Thanos' reply {\textcolor{blue}{[In order for $\mathcal{I}_{\mathrm{dS}}=0$, we have to set  $\psi _0 ^{(\mathrm{dS})} = -4+\ln 4$. You can check that eq.(22)  matches eq.(32) when $M=0$ whatever the value of $\psi _0 ^{(\mathrm{dS})}$ and $\psi _0 ^{(\mathrm{L})}$, 	as long as $\psi _0 ^{(\mathrm{dS})}=\psi _0 ^{(\mathrm{L})}$.]}}}  

\section{Schwarzschild-de Sitter instanton}
\label{sds_instanton}

\noindent For the Schwarzshild-de Sitter spacetime, the solution of the field equations \eqref{a9} can be derived after imposing reflection symmetry $(x \rightarrow |x|)$ around the origin. The line element then becomes \cite{MST90}
\begin{equation} \label{c1}
\d s^2=-(C+2 M |x| - |\Lambda| x^2) \d t^2 + \frac{\d x^2}{C+2 M |x| - |\Lambda| x^2} \, .
\end{equation}

\noindent The signs of $M$ and $C$  determine several distinct classes of solutions and their 
%relative 
associated causal structures \cite{MuN11}. Here we require $M>0 \,$ because a negative mass implies a naked singularity \cite{MST90}.
%% \textbf{[PN: Thanos, I do not understand this statement. For a naked singularity you need to have a horizonless geometry, i.e., a negative discriminant $M^2 + C|\Lambda|<0$]. As you can see the sign of $M$ plays no role. It is only the sign of $C$ that can lead to naked singularities. Can you check this please?]}. 
Moreover, the existence of a compact spacetime, having both   an event horizon $x_{\mathrm{h}} \,$ and a cosmological horizon $x_{\mathrm{c}} $, is possible only if \ $M^2 + C|\Lambda| \geq 0 \,$ with  $C<0 \,$. 
%% \textcolor{red}{[But $C >0$ in background?]} 
Since the value of $C$ is arbitrary, we can put $C=-1 \,$. 
%\noindent \textcolor{red}{ [But $C = +1$ in 3D Schwarzschild, so give 1D Schwarzschild explicitly to clarify.]} \textcolor{blue}{[This issue has been discussed in Piero's latest email. For $C>0$ we get a naked singularity at the origin and so we cannot build a regular instanton. That's the reason we discard this case.]} 
On symmetry grounds, we consider the spacetime region with $x>0 \,$ and drop the absolute value 
%for the construction of 
in the expression for the instanton. The general form of the potential for the $(1+1)$-dimensional  Schwarzschild-de Sitter instanton is therefore
\begin{equation} \label{c2}
V(x)= -1+2 M x - |\Lambda| x^2 \quad \mathrm{for} \quad x>0,
\end{equation} 
%
%% \noindent \textcolor{red}{ [This should match eqn (14) when $M=0$, so signs may be wrong]} \textcolor{blue}{[The correct answer is $V(x)= -1+2 M x - |\Lambda| x^2 \,$.]} 
with the corresponding  horizons
\begin{eqnarray}
x_{\mathrm{h}}&=&\frac{1}{|\Lambda|} \left( M-\sqrt{M^2 - |\Lambda|}\right)  \equiv a \label{c3} \\
x_{\mathrm{c}} &=& \frac{1}{|\Lambda|} \left( M+\sqrt{M^2 - |\Lambda|}\right) \equiv b \, .  \label{c4}
\end{eqnarray} 
%
%% \noindent  \textcolor{red}{[$C= +1$ gives $\sqrt{M^2 + |\Lambda |}$, so $x_h < 0$.]} 
One gets two distinct horizons for $M^2 > |\Lambda| \,$  (the \textit{lukewarm} case) but a degenerate horizon for $M^2 = |\Lambda| \,$ (the \textit{Nariai} case). These two cases lead to smooth and regular  Euclidean manifolds. 
%Remarkably, in both cases the two horizons have a common Hawking temperature and that makes our spacetime classically and quantum-mechanically stable \textbf{[PN: I do not understand this sentence. DeSitter will be always unstable quantum mechanically]}. 

\par   Let us start with the lukewarm case
%$(1+1)$-dimensional lukewarm instanton 
\cite{Rom92,CJS98}. The event horizon
%, $x_{\mathrm{h}}=a \,$, 
is given by \eqref{c3} and the cosmological horizon
%, $x_{\mathrm{c}}=b \,$, is given 
by \eqref{c4}. We analytically continue to the Euclidean section by putting $\tau=it \,$ and then choose the region $a \leq x \leq b \,$ where the metric is positive-definite. Remarkably the two horizons have the same temperature. Therefore we can remove the conical singularities by demanding $\tau$ to be periodic with the period being the inverse of the
% usual \textcolor{red}{[BUT IT'S NOT!]} Hawking 
temperature
% [CONFUSING TO CALL THIS HAWKING TEMPERATURE]:
\begin{equation} \label{c5}
\beta=T^{-1}=\left( \frac{1}{2 \pi}\sqrt{M^2-|\Lambda|}\right) ^{-1} \, .
\end{equation} 

\noindent As expected, the 
%Hawking 
temperature  in $(1+1)$-dimensions is proportional to the black hole mass \cite{Mur12} for $M \gg \sqrt{\Lambda}$, so heavier black holes are hotter.
%, while lighter black holes are colder. 
This means they
%black holes 
have a positive heat capacity and 
%the evaporation will 
relax towards smaller, colder configurations. 

\begin{figure}[h!] 
\begin{center}
\includegraphics[width=0.7 \textwidth]{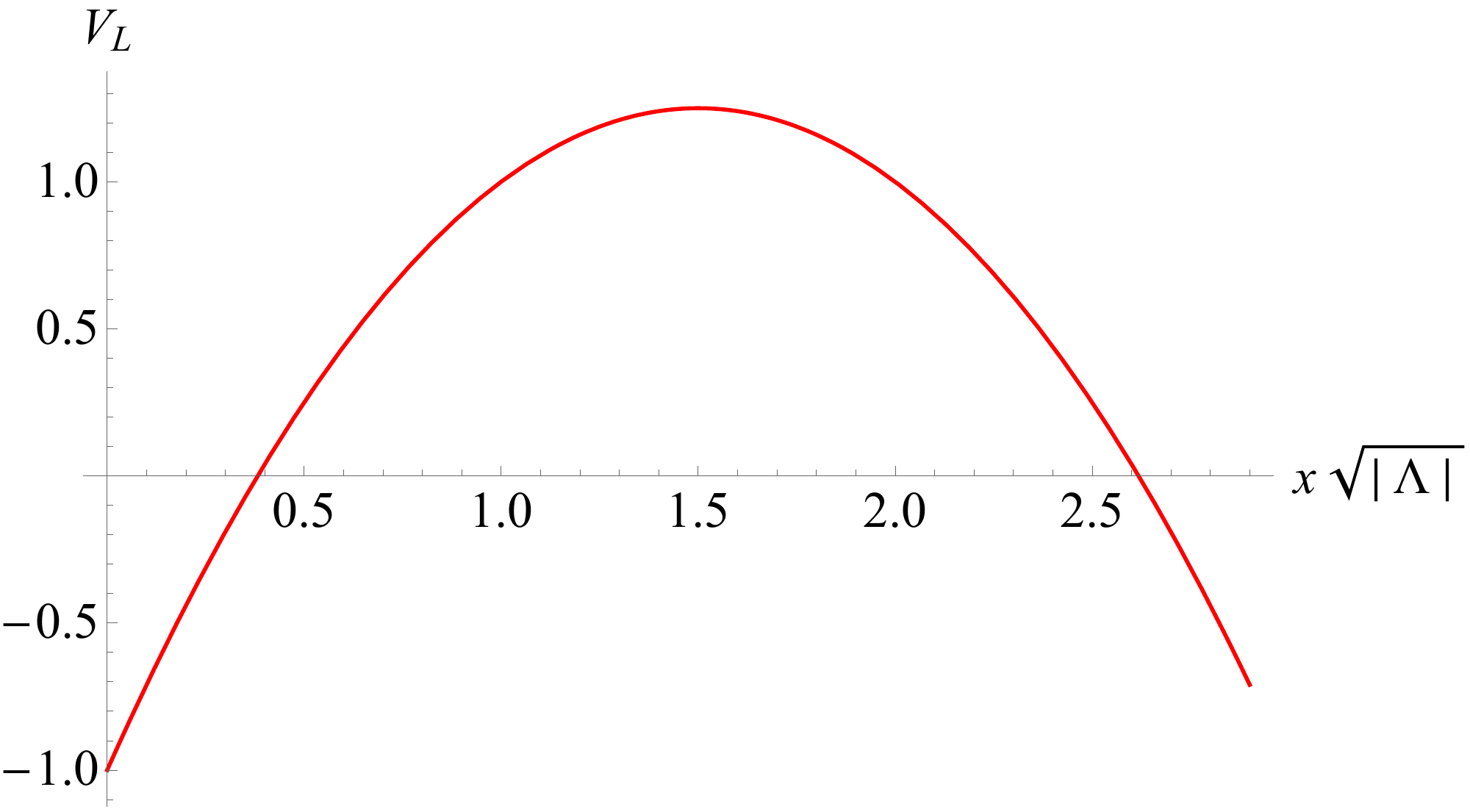}
\caption{The function $V_{\mathrm{L}}(x)\,$
\textit{vs} $x\sqrt{|\Lambda|}$ for $M > \sqrt{|\Lambda|}$. }
\label{fig:Fig. 2}
\end{center}
\end{figure}

One can write the metric coefficient
%metric coefficient for
in the lukewarm case as
\begin{equation} \label{c6}
V_{\mathrm{L}}(x)=\frac{(x-a)(b-x)}{ab} \, ,
\end{equation} 

\noindent whose  form
%topology 
is plotted  in Fig. \ref{fig:Fig. 2}. 
%%\textbf{[PN: Thanos, here again we do not need an additional scale $\ell$. Only the relation $M$ vs $\Lambda$ counts.]}
%\noindent 
The   instanton action for the Schwarzschild-de Sitter spacetime  is  
\begin{equation} \label{c7}
\mathcal{I}_{\mathrm{SdS}}=  - \frac{|\Lambda|}{8 \pi G_{(2)} } \int \d^2x \sqrt{g} \     \left(   \frac{1}{2} \psi  +1  \right) \, 
\end{equation} 

\noindent and from \eqref{a6} one obtains 
%for the dilaton $\psi$ 
\begin{equation} \label{c8}
\psi = \psi_0^{(\mathrm{L})} -  \ln \left( -1+2Mx-|\Lambda| x^2 \right), 
\end{equation} 
where $\psi_0^{(\mathrm{L})}$ is an integration constant.
 The  lukewarm instanton action is therefore
\begin{equation} \label{c9}
\mathcal{I}_{\mathrm{L}}  =  -\frac{1}{8G_{(2)}} \left[ 2\psi_0^{(\mathrm{L})}+ 8 + \ln \left( 
\frac{\Lambda^2}{16 (|\Lambda|-M^2)^2}  \right)\right] \,
\end{equation} 
%\textcolor{red}{If $\psi_0^{(\mathrm{L})} = \psi_0^{(\mathrm{dS})}$, this agrees with eqn (22) for $M=0$, whatever the value of  $\psi_0^{(\mathrm{L})}$}, \textcolor{blue}{[Thanos: Even better, we can set $\psi_0^{(\mathrm{L})}=-4 + \ln 4$ to eliminate also the 16 in the denominator of eq.(32), as also mentioned in Sec. 2 ]} \textbf{but $\psi_0^{(\mathrm{L})}$ can be set to $-4$ to eliminate terms that do not depend on the scales of the system. [PN: please Thanos check is this is consistent.]} 
%Accordingly the lukewarm instanton-action finally reads}
%% \textbf{It therefore becomes}
%% \begin{equation} \label{c9bis}
%% \mathcal{I}_{\mathrm{L}}  =  -\frac{1}{8G_{(2)}} \ln \left( \frac{\Lambda^2}{16 (|\Lambda|-M^2)^2}  \right) \quad \mathrm{for} \quad M \textcolor{red}{<}\sqrt{|\Lambda|} \, \textcolor{red}{[OK?]} \, .
%% \end{equation} 
%% 
%% \textcolor{blue}{[The corrected form of eq.(32) is:
and this becomes 
\begin{equation} \label{c9bis}
\mathcal{I}_{\mathrm{L}}  =  -\frac{1}{8G_{(2)}} \ln \left( \frac{\Lambda^2}{ (|\Lambda|-M^2)^2}  \right) \quad \mathrm{for} \quad M > \sqrt{|\Lambda|}
\end{equation} 
%for
providing we choose
\begin{equation}
\psi_0^{(\mathrm{L})}=\psi_0^{(\mathrm{dS})}=-4 + \ln 4 \, .
%% ]}
\end{equation}

For the  $(1+1)$-dimensional Nariai instanton, $M^2=|\Lambda| \,$ and so  there is a degenerate horizon at 
\begin{equation} \label{c10}
x_{\mathrm{h}}=x_{\mathrm{c}}=\rho=\frac{M}{|\Lambda|} \, .
\end{equation} 
	
\noindent   The double root implies that the proper distance from any point to the degenerate horizon is infinite \cite{MaR95}. 
%there is no 
The surface gravity on the horizon is therefore zero,
%$(\kappa=0) \,$, 
%since it 
corresponding to the extremal case, and 
%one can, however, express 
the mass and cosmological constant are
%with respect to the double root as
\begin{eqnarray}
M=\frac{1}{\rho} , \quad
% \label{c11} \\
|\Lambda| = \frac{1}{\rho^2}  \label{c12} \, .
\end{eqnarray} 

\begin{figure}[h!]
\begin{center}
\includegraphics[width=0.7 \textwidth]{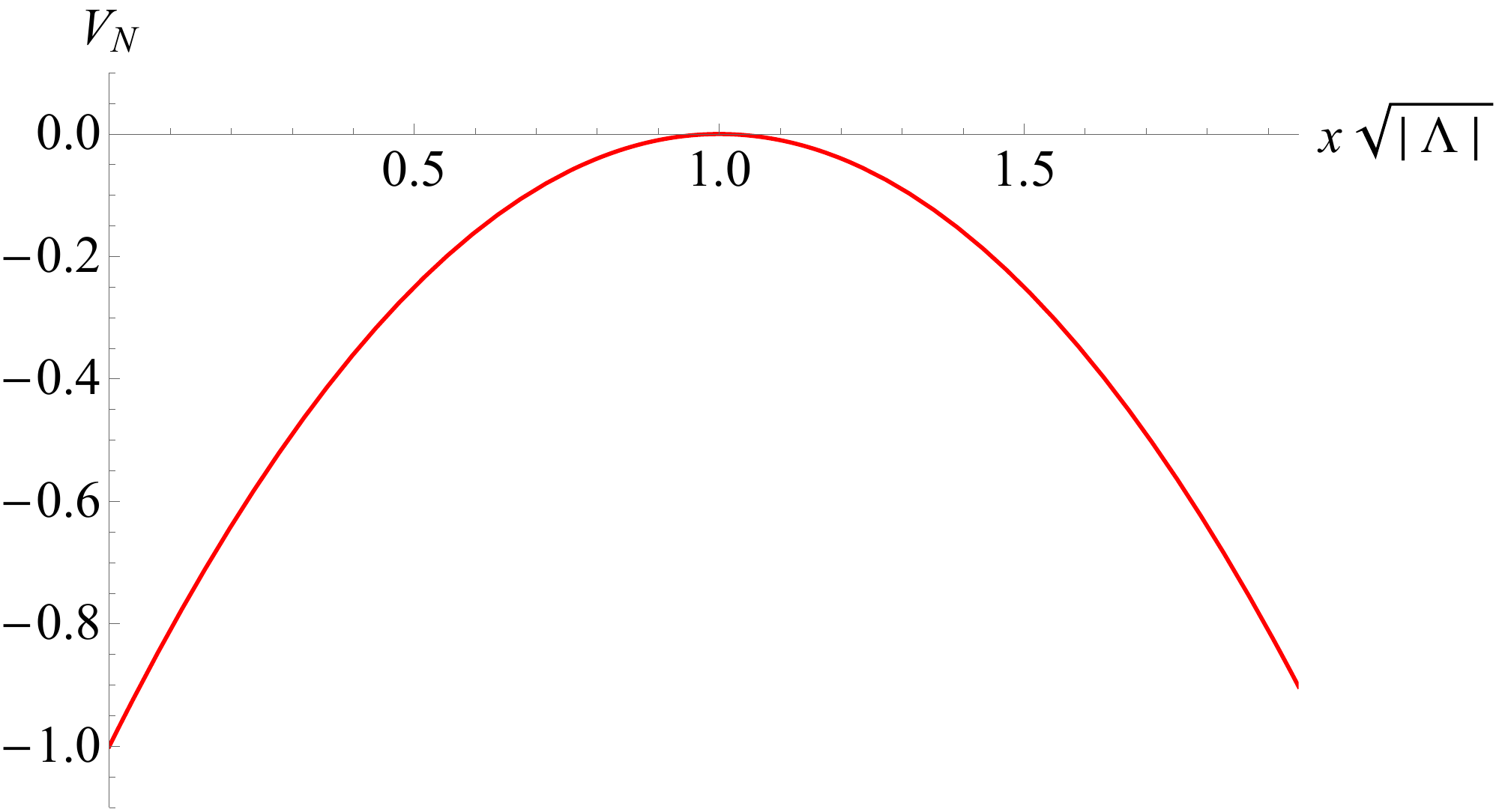}
\caption{The function $V_{\mathrm{N}}(x)\,$
\textit{vs} $x\sqrt{|\Lambda|}$ for $ M = {\sqrt{|\Lambda|}}$. }
\label{fig:Fig. 3}
\end{center}
\end{figure}

\noindent 
%[EQNS MERGED] 
By substituting \eqref{c12} 
%and \eqref{c12} 
into \eqref{c2}, one therefore gets the form of the potential for the $(1+1)$-dimensional Nariai instanton,

\begin{equation} \label{c13}
V_{\mathrm{N}}(x)=-\frac{(x-\rho)^2}{\rho^2}  \, ,
\end{equation} 

\noindent and this is illustrated  in Fig. \ref{fig:Fig. 3}. 
%%\textbf{[PN: Thanos again get rid of unnecessary scales]} 

%\noindent The form of $V_{\mathrm{N}} \,$ 
By Wick-rotating $(\tau=it) \,$ we get the regular instanton metric
\begin{equation} \label{c14}
\d s^2= -\frac{(x-\rho)^2}{\rho^2}  \d\tau^2 - \frac{\rho^2 }{(x-\rho)^2} \d x^2 .
\end{equation}

\noindent To demonstrate
%see better 
its regularity,
 %of this instanton, 
we can apply the 
%proposed 
Nariai transformation \cite{HaR95,Bou97}: 
\begin{eqnarray}
\tau = \frac{\xi}{|\Lambda| \epsilon}, \quad
% \label{c15} \\
x= \rho - \epsilon \cos\chi
, \quad
% \\  \label{c16}
x_{\mathrm{h}} =\rho - \epsilon ,\quad
% \\ \label{c17}
x_{\mathrm{c}} = \rho + \epsilon \, ,
%  \label{c18}
\end{eqnarray} 

\noindent
% [EQNS MERGED]
 where the two horizons coincide in the limit $\epsilon \rightarrow 0 \,$. 
 Under this transformation, the instanton \eqref{c14} can be written  in the form of \eqref{b9}, 
%\begin{equation} \label{c19}
%\d s^2=\frac{1}{|\Lambda|}\left(  \d\chi^2 +  \sin^2\chi \ \d\xi^2 \right) 
%\end{equation} \vspace{0.1mm}
%\noindent 
%with  $\xi \in [0,2\pi] \,$ and $\chi \in [0, \pi]$. 
%[EQUATION REMOVED SINCE SAME AS (19).]
so this can again be transformed into the inflating form (\ref{eq:b9}). 
One might wonder why this does not represent 
%As shown in the Appendix, this can be transformed into 
a black hole in an expanding background; the reason is that the black hole and cosmological event horizons coincide in this case, so -- in some sense -- the background universe is itself a black hole. 
%SURELY THIS IS OBVIOUS SINCE IT'S THE SAME METRIC].
%\textbf{[PN: I think these domains are incorrect since they imply unwanted degeneracy $0, 2\pi$. Thanos can you check it please].}  \textcolor{red}{This agrees with eqn (18), as expected.}
%[CAN THIS BE TRANSFORMED INTO A BLACK HOLE IN AN EXPANDING RW BACKGROUND (APPENDIX)?]

%Finally the calculation of 
For the Nariai instanton ($M=\sqrt{|\Lambda|}$), the action turns to be
\begin{equation} \label{c22}
\mathcal{I}_{{\mathrm{N}}} = - \frac{\psi_0^{({\mathrm{N}})}+4-\ln\left(4 \epsilon^2 |\Lambda|  \right)}{4 G_{(2)}} \, ,
\end{equation} 
where $\psi_0^{({\mathrm{N}})}$ is an integration constant.
%% which can  be chosen \textcolor{red}{to be $-4 + 2 \ln \epsilon$}  to eliminate the dependence on $\epsilon$. 
%As a result the Nariai instanton-action turns to be
The action then becomes
%}
\begin{equation} \label{c22bis}
\mathcal{I}_{{\mathrm{N}}} =  \frac{\ln\left(|\Lambda|/\mu_0^2  \right)}{4 G_{(2)}} 
% \quad \mathrm{for} \ M=\sqrt{|\Lambda|}
\end{equation} 
providing we choose 
\begin{equation}
\psi_0^{({\mathrm{N}})}= -4 + 2 \ln(2 \epsilon \mu_0) \, .  
\end{equation}

\noindent Here $\mu_0$ is an arbitrary mass scale, at least in principle
%, is not 
unconnected to $M$ or $\mpl$.

\section{Creation rate of PBHs in $(1+1)$-dimensions}
\label{2d_rate}

%Primordial black holes are theoretical astrophysical objects that  are expected to exist  in the very first moments after the Bing Bang, where the conditions of the universe were violent. They play a critical role in a variety of early-universe processes \cite{Car77,CaS83,UGS99} and are also possible dark matter candidates \cite{FKT10,CKS++10,CKS16}. Therefore, understanding their evolution in a dimensionally-reduced universe can shed new light on modern Cosmology. Furthermore, there have been proposed possible theories for their formation, such as the collapse of primordial inhomogeneities \cite{Car05}, or cosmological phase transitions \cite{Car03}, but here we are interested in the quantum   formation of black holes out of nothing, due to the strong expansion of the universe. 

We now have  all the elements needed to calculate the rate  -- or, more strictly, probability --  of PBH production in $(1+1)$-dimensions.
Before going into the details, it is worth recalling what has been obtained so far.
First, 
the ``rate'' depends only on the instanton action of the nucleated object,
\begin{equation} 
\Gamma=\exp\left[ -2\mathcal{I}_{\mathrm{bh}} \right] \, ,
\end{equation} 
so the de Sitter background does not contribute to the rate, which is a great simplification.
% since the production rate 
Second, the two-dimensional topology allows for a richer instanton structure  than
%t is missing
 in four dimensions (eg. the lukewarm case). Third, the absence of a Planck mass in two dimensions has the important consequence that the de Sitter space is unstable, irrespective of the values of the cosmological constant and the black hole mass. Only  their ratio affects the decay rates. This 
%fact largely 
extends the $(3+1)$-dimensional scenario 
%of the $(3+1)$-spacetime 
and allows for sub-Planckian PBH nucleation.

\begin{figure}[h!]
\begin{center}
\includegraphics[width=0.75 \textwidth]{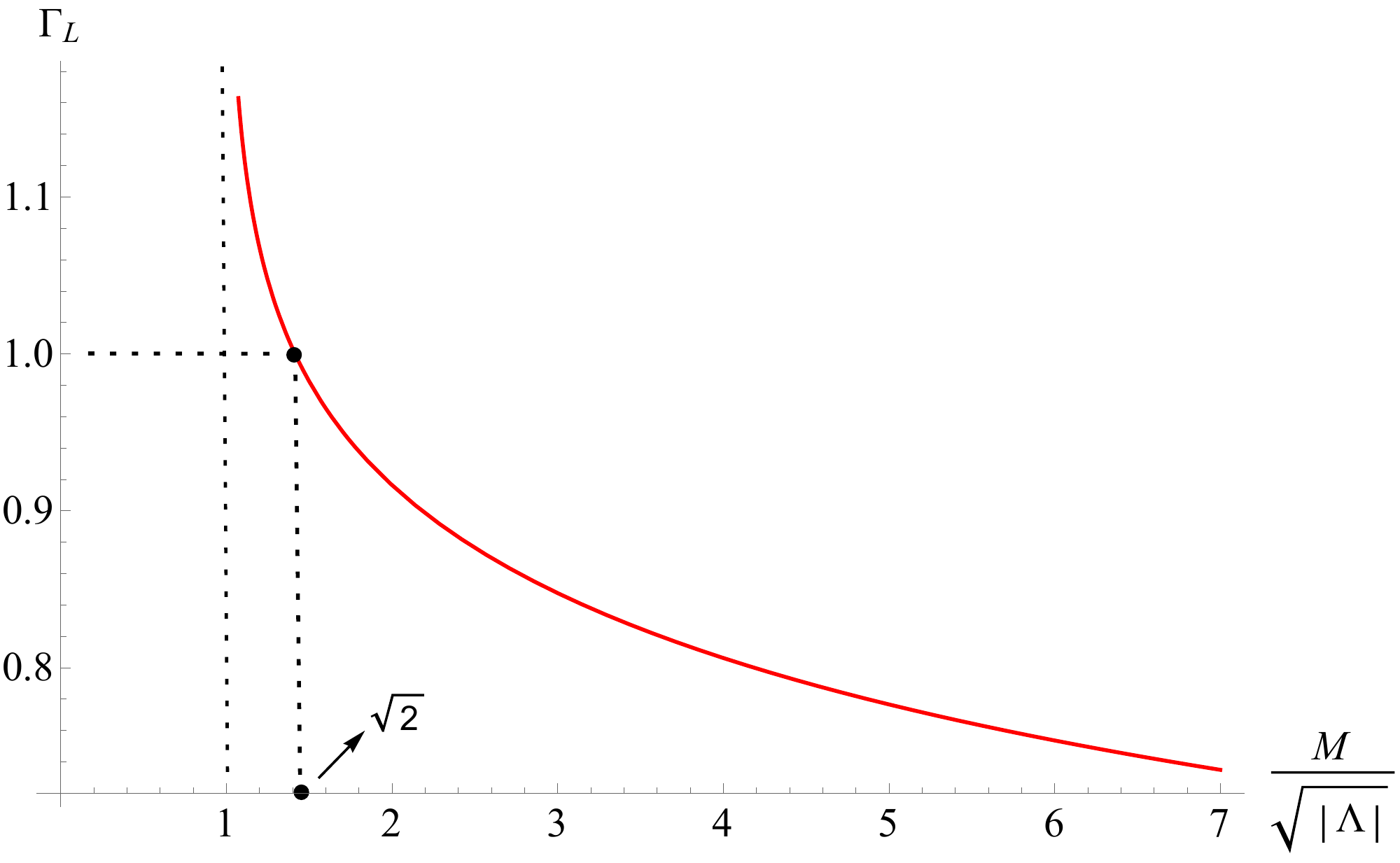}
\caption{The rate $\Gamma_{\mathrm{L}}$ \textit{vs} $M/\sqrt{|\Lambda|}$ for $M > \sqrt{|\Lambda|}\,$.}
\label{fig:Fig. 4}
\end{center}
\end{figure}

We now focus on the 
%first case, namely the
 $(1+1)$-dimensional lukewarm spacetime, for which the rate is
%. The resulting rate reads
\begin{equation} \label{d1}
\Gamma_{\mathrm{L}}=  \left[ \frac{\Lambda^2}{( M^2 - |\Lambda|)^2} \right]^{\frac{1}{4G_{(2)}}} \, \, .
\end{equation} 
%% \textcolor{blue}{[The factor of 16 does not affect the result but if we follow the new proposed values for the boundaries, then the correct form of the rate is:
%% \begin{equation} \nonumber
%% \Gamma_{\mathrm{L}}= \left[ \frac{\Lambda^2}{( M^2 - |\Lambda|)^2} \right]^{\frac{1}{4G_{(2)}}}\, 
%% \end{equation} 
%% ]}
%%
%\noindent \textcolor{red}{[$\Gamma_L$ later? No factor of $16$ in 2nd expression?]} where we have used
% taken into account that 
%$G_{(2)}=2\pi \,$ in natural units. \textbf{[PN: Thanos you have to fix the exponent here, I think. Probably we do not need to use $G_{(2)}=2\pi \,$. Please express it and we decide later what is the most convenient value to assign it.]} \textcolor{red}{[$G_{(2)}=1$ seems more sensible but does extra factor of $2$ matter?]}
We plot this as a function of
%the above rate with respect to the 
mass for a fixed $|\Lambda| \,$ in Fig. \ref{fig:Fig. 4}. We see that the rate increases as the mass of the black hole decreases.
To summarise, one has:
\begin{itemize}
\item For $M\gg\sqrt{|\Lambda|} \,$, the rate is highly suppressed $(\Gamma_{\mathrm{L}} \ll 1) \,$.
\item For $M=\sqrt{2 |\Lambda|} \,$,   $\Gamma_{\mathrm{L}}=1 \,$ and the two universes have equal probability.
%($P_{\mathrm{dS}} = P_{\mathrm{L}} \,$).
\item   For  $\sqrt{|\Lambda|} < M < \sqrt{2|\Lambda|} \,$, the rate $\Gamma_{\mathrm{L}}$ exceeds $1$, %($\Gamma_{\mathrm{L}} > 1 \,$), 
corresponding to a highly unstable de Sitter space.  
\item For $M \approx \sqrt{|\Lambda|} \,$, the rate diverges ($\Gamma_{\mathrm{L}} \gg 1$).
\end{itemize}

%More specifically, if $M=\sqrt{2 |\Lambda|} \,$ then  $\Gamma_{\mathrm{L}}=1 \,$. That means, the two universes have equal probabilities, $P_{\mathrm{deSitter}} = P_{\mathrm{lukewarm}} \,$, since the two instanton actions are equal at this value, \ $\mathcal{I}_{\mathrm{dS}}=\mathcal{I}_{\mathrm{l}} \,$. 
%If  $M > \sqrt{2 |\Lambda|} \,$, the rate is suppressed having a range of  $0 < \Gamma_{\mathrm{L}} <1$. So, the formation of PBHs with masses close to the value $M \sim \sqrt{2|\Lambda|}$ is favoured over the heavier ones. On the contrary, if the mass range is $\sqrt{|\Lambda|} < M < \sqrt{2|\Lambda|} \,$, the rate becomes unsuppressed, $\Gamma_{\mathrm{L}} > 1 \,$, and so de Sitter space would be highly unstable to the formation of black holes.  
%Furthermore, if $\Lambda \rightarrow 0 \,$, the rate also drops to zero,  meaning that, if the universe is strongly expanding, the spontaneous black hole formation becomes more considerable. 

\begin{figure}[h!]
\begin{center}
\includegraphics[width=0.7 \textwidth]{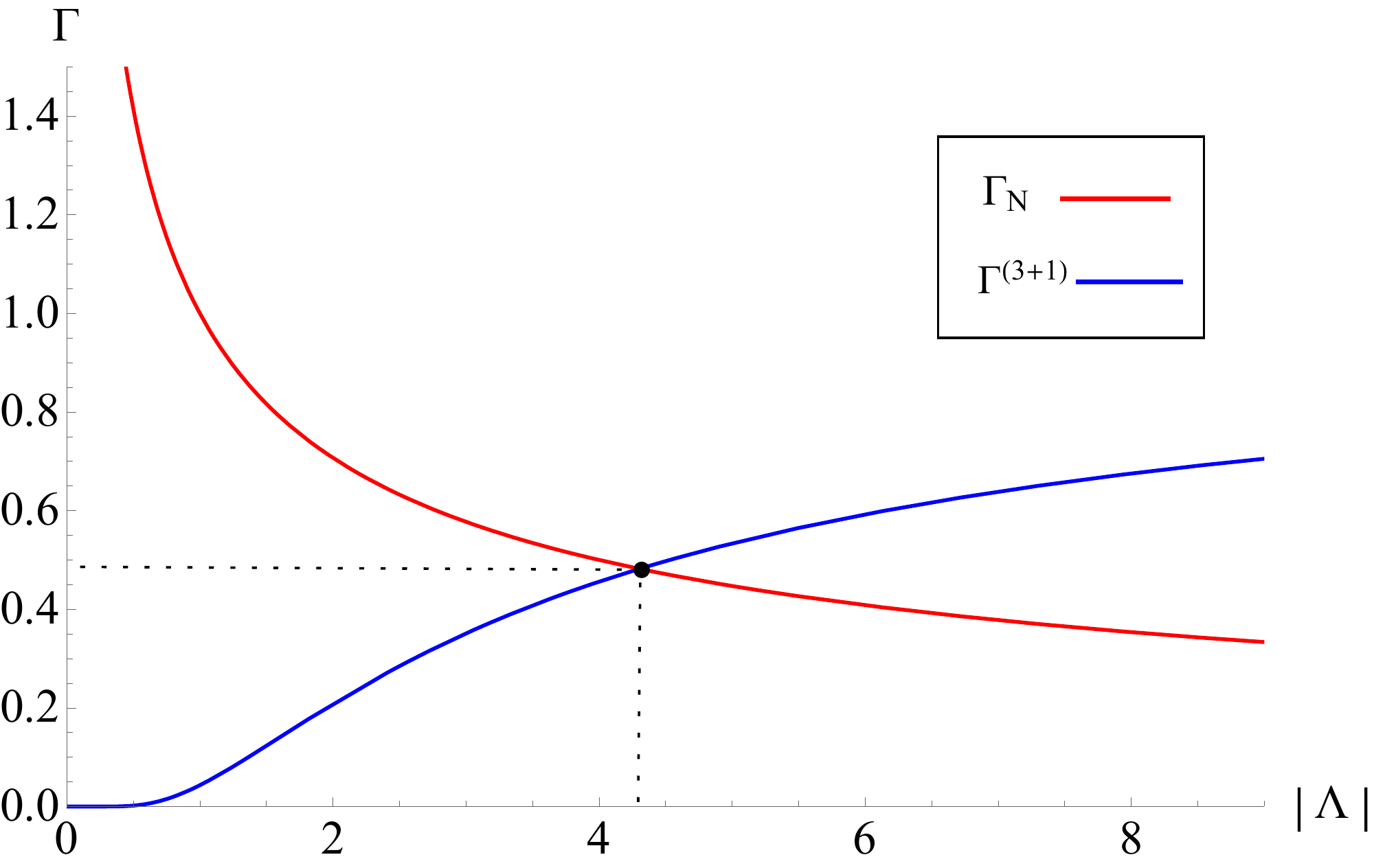}
\caption{Rate comparison: $\Gamma_{\mathrm{N}}$ and $\Gamma^{3+1}$ \textit{vs} $|\Lambda|=\Lambda_\mathrm{c}$ for $\mu_0 = G_\mathrm{N}=G_{(2)}= 1 \,$. }
\label{fig:Fig. 4 vs 2}
\end{center}
\end{figure}

Next we consider
%shift our interest to 
the $(1+1)$-dimensional Nariai instanton. 
% by substituting  $\mathcal{I}_{\mathrm{bh}}=\mathcal{I}_{\mathrm{N}} \,$ into the expression of the rate \eqref{a3}. At this case we imply different boundary conditions for the two spacetimes, otherwise we will get a divergent rate. Thus, we can write the two instantons as
%\begin{eqnarray}
%\mathcal{I}_{\mathrm{dS}} &=& - \frac{\psi _0 ^{(\mathrm{dS})} +4 - \ln 4}{4G_{(2)}} \\
%\mathcal{I}_{\mathrm{N}} &=& - \frac{\psi _0^{(\mathrm{N})} +4 - \ln (4 \epsilon ^2 |\Lambda |)}{4G_{(2)}}
%\end{eqnarray}
%We choose $\psi _0 ^{(\mathrm{dS})}=0\,$, while for the Nariai-boundary $\psi _0^{(\mathrm{N})} = \ln (4\mu _0 ^2 \epsilon^2)\,,$
%where $\mu_0$ is a minimum cut-off mass. That way the Nariai instanton reads
%\begin{equation}
%\mathcal{I}_{\mathrm{N}} = - \frac{ 4 - \ln \left( \frac{|\Lambda |}{\mu _0 ^2} \right) }{4G_{(2)}}
%\end{equation}
%and the Nariai rate will be
After choosing suitable boundary conditions, one obtains
%\begin{equation}
%\Gamma_{\mathrm{N}} = \left( \frac{4 \mu _0 ^2}{|\Lambda |} \right)^{\frac{1}{4\pi}}  \, .
%\end{equation}
%\textcolor{blue}{[In $\Gamma_{\mathrm{N}}$ there should be no factor of 4. The correct form is:
\begin{equation} %\nonumber
\Gamma_{\mathrm{N}} = \left( \frac{ \mu _0 ^2}{|\Lambda |} \right)^{\frac{1}{2G_{(2)}}}  \, .
\label{nariairate}
\end{equation}
%]}
Here $\mu_0$ is not set {\it a priori} and PBHs can be produced prolifically for any value of $\Lambda$. As shown in Fig. \ref{fig:Fig. 4 vs 2}, the Nariai instanton allows comparison with the four-dimensional case.
%For this purpose, we set $\mu_0\sim \mpl$ and %we consider 
%the ratio $\Gamma_{\mathrm{N}}/\Gamma^{3+1}$ is then as shown in Fig. \ref{fig:Fig. 4 vs 2}.

%One can see that 
%\textcolor{red}{[SOME REPHRASING]} 
 We now consider models in which the  2D and 4D cosmological constants are equal (i.e. $| \Lambda |=\Lambda_\mathrm{c}$). From eqs.~(\ref{a4}) and (\ref{nariairate}), the two-dimensional PBH production rate then dominates the four-dimensional one for any value of the (joint) cosmological constant 
%[WHICH ONE?] 
smaller than a critical value:
\begin{equation}
%|\Lambda|=
\Lambda_\mathrm{c} < 
\Lambda_\mathrm{crit} \simeq 4.31 \mu_0^2\, .
\end{equation}
%\textbf{[PN: Thanos please find the value at which the two rates coincides. Maybe plot the curve longer on the right]}. 
%[REPHRASED AS REQUESTED BUT I UNDERSTAND NEITHER FORM NOR MEANING OF THIS EQUATION.
%COMPARING (4) AND (47) DOES NOT GIVE (48).  ALSO EQUAL PRODUCTION RATES DOES NOT IMPLY EQUAL COSMOLOGICAL CONSTANTS IF $G_{(2)} \neq G_N$.]
 In general, the cosmological terms $\Lambda$ and $\Lambda_\mathrm{c}$, entering the two-dimensional and four-dimensional gravity actions, are not related. Even if action \eqref{a5} is the dimensionally reduced 
%version of the 
Einstein-Hilbert action, there is no reference scale to set  $\Lambda$, since $G_{(2)}$ is dimensionless in two dimensions. We have already noted such a scale freedom, since the production rate depends on  either $M^2/\Lambda$ or $\mu_0^2/\Lambda$ but not $\Lambda$ itself. As a result, due to the arbitrariness of $\Lambda$, we can have two additional regimes:
%, namely 
$|\Lambda|/\Lambda_\mathrm{c}\ll 1$ and
$|\Lambda|/\Lambda_\mathrm{c}\gg 1$.
%THIS IS NOT CLEAR]

The first regime implies a prolific production of light two-dimensional Nariai black holes with
$M=\sqrt{|\Lambda|}$. For ease of discussion, one can describe these as
%speak of 
black holes with sub-Planckian masses. To this end, we indirectly introduce Planck units in two dimensions via the relation between   $\Lambda$ and $\Lambda_\mathrm{c}$, \textit{e.g.} $\Lambda\ll \Lambda_\mathrm{c}\sim 1$ for $G_{\mathrm{N}}=1$. Such sub-Planckian black holes have large radii and are thermodynamically stable due to their 
%vanishing temperature and 
positive heat capacity.
 %[NOT VANISHING TEMPERATURE]. 
Eventually they
%such black holes 
oxidate, \textit{i.e.}, undergo a phase transition to the four-dimensional form.
%Universe. 
Unfortunately, the Schwarzschild metric cannot be used to connect a $1+1$ model with a $3+1$ one. The major obstacle is the presence of a $2M/r$ term, which contrasts with the $Mr$ term which appears 
%that is in contrast with the power of the spatial coordinate 
in two dimensions.
%, namely $\sim Mr$. 
Rather one should employ the recently proposed holographic metric \cite{NiS14,FKN16}  
\begin{equation}
\d s^2= -\left(\,1-\frac{2M\LP^{2}\, r}{ r^2
+L^2} \,\right)dt^{2} +\left( \, 1-\frac{2M\LP^{2}\, r}{ r^2+L^2  } \, \right)^{-1}dr^{2}+r^{2}d\Omega^{2},
\label{eq:holo}
\end{equation}
with $L\sim\LP=1/\Mpl$. 
%One can notice that 
For $r\gg L$, 
%\eqref{eq:holo} 
this becomes the Schwarzschild spacetime, while for  $r\ll L$ it becomes 
%one finds an 
the effectively two-dimensional metric
\begin{eqnarray}
\d s^2\longrightarrow &-\left(\,1-2M r \LP^2 /L^2 \,\right)dt^{2}
+\left( \, 1-2M r \LP^2 /L^2\, \right)^{-1}dr^{2}+\mathcal{ O}\left(r^2/L^2\right).
\end{eqnarray}
%\PN{The next sentence should be moved in Sec. 5 because where we clarify that the 1+1 phase is an effective one, while the spacetime is still 3+1 and allows for the definition of sub-Planckian masses. Otherwise the adjective sub-Planckian does not make sense in two dimensions.} In particular, one finds prolific production of two-dimensional Nariai sub-Planckian PBHs with $M=\sqrt{|\Lambda|}$. These have large radii and are thermodynamically stable due to their low \textcolor{red}{(zero?)} temperature.  \PN{yes, actually zero ... but because the heat capacity is positive there!}
% [MISSING FACTOR OF $(L_P/L)^2$ IN METRIC?]

The second regime, $|\Lambda|/\Lambda_\mathrm{c}\gg 1$,  corresponds to a two-dimensionally driven inflationary scenario that dominates over the standard four-dimensional one. The expansion is now
%would be 
controlled by the scale factor appearing in \eqref{eq:b9}. 
%In \cite{ChM95} 
Chan and Mann argue that a pure-radiation model is static and that one needs some exotic matter content to drive inflation \cite{ChM95}.  However, their framework is different from ours, since they do not include a cosmological term in the gravity action. In our model, the presence of a negative  $\Lambda$ suffices.  
%A full discussion of how $1+1$ models can expand exponentially for large $\Lambda$ is beyond the scope of the current paper.
% [BUT THE PAPER IS  OBSCURE WITHOUT THIS CLARIFICATION.]
%\PN{Sentences taken from Sec. 5}

\section{Observational consequences}
\label{observations}

%[ADAPTED FROM FROM PIERO'S EMAIL] 
In order to understand the observational consequences of this scenario, we need to consider the implications of the $1+1$ phase for the standard $3+1$ inflationary model. If we follow the Bousso-Hawking and Mann-Ross approaches, which assume  $3+1$ dimensions,
the quantum mechanical decay of the universe into a
spacetime with a black hole occurs only when the cosmological constant has 
the Planckian value of around  $10^{133}$ eV m$^{-3}$.
 In the usual scenario, $\Lambda_\mathrm{c}$ has an energy density of $10^{117}$ eV m$^{-3}$ at the
start of inflation, which is $10^{16}$  times smaller,
% than this, $M_p/l_p^3$would correspond to 
so  
%This means that $\Lambda$ in geometricalunits is close to $10^{-8}$, which is rather far from $1$. 
%The conclusion is that this value of the cosmological constant at the beginning of the inflation might, 
there should be no black hole production by this mechanism in $3+1$ dimensions.
%according to the Bousso-Hawking and Mann-Ross mechanisms. 
However, in $1+1$ dimensions, there is no Planck scale, since Newton's constant is
dimensionless, so black holes can be produced even without Planckian values of the cosmological constant. 
%This  motivates our present considerations. 
Indeed, dimensional reduction in quantum gravity may occur at an energy far below the Planck scale. 
%rrespective of whether the spacetime dimension is $3+1$ or $1+1$. 
%Only if the Planckian phase for the cosmological constant persists throughout inflation can this conclusion be avoided. 
\if
Atick and Witten showed that a gas of strings, heated up to a critical temperature, $T_\mathrm{crit}$, below the Hagedorn temperature, undergoes a phase transition to an effective two-dimensional phase \cite{AtW88}. 
% \PN{The last sentence has been added in support of the persistence of the 1+1 phase. We mentioned it also in the caption of the new Fig. 1}. We therefore consider  the possibility that the $1+1$ phase persists throughout inflation. 
%In this case, we must justify the compatibility the $1+1$ inflationary scenario with observations. 
In this context, one  might {\it expect} the $1+1$ phase to be inflationary %\textcolor{red}{[CHECK]}  
 since one can presumably choose a spacelike slicing in which the metric \eqref{b9} has an exponentially expanding form \eqref{eq:b9} as with the $3+1$ de Sitter solution. 
%\PN{We can leave out the word ``presumably'' since we showed this.}. 
The issue of whether $1+1$ models can expand exponentially has been studied by Chan and Mann \cite{ChM95}. They argue that a pure-radiation model is static and that one needs some exotic matter content to drive inflation. However, there is no such content in our model %[?]. \PN{yes} 
%there is a compromise which avoids invoking a new inflationary scenario. 
\fi

If the $1+1$ phase
% for the cosmological constant (phase when its value is close to $1$ in geometrical units) 
ends  before inflation is complete, there are  no observational consequences because the black hole number density is exponentially diluted. However, we stress that we are only considering
``effective'' $1+1$ dimensionality. This is because `t Hooft's argument
for the self-renormalizability of gravity requires that
gravity perceives just $1+1$ dimensions but
 the Universe itself must remain $3+1$ dimensional 
%Although it is $3+1$ dimensional,only $1+1$ dimensions are felt or 
%perceived or actually playing a role.Note that there is 
%one still needs a $3+1$ ambient space to 
in order to describe the
ultraviolet (Planckian) regime. 
%where gravity actually behaves $1+1$ dimensionally. 
Otherwise there would be no Planck scale at all. For present  purposes, we therefore adopt a compromise model, in which the $1+1$ phase persists throughout 
inflation, while the
ambient cosmological space evolves $3+1$ dimensionally, as in the standard
model. In this way, we avoid modifying the  standard inflation scenario but still 
expect observational consequences from PBH production during the $1+1$ phase.
%In contrast to the $3+1$ case, PBH production rates  do not need Planckian values of the cosmological constant since there is no scale of reference coming from the Newton's constant. 
%SO DO WE HAVE $a \propto \exp(\sqrt{|\Lambda |} \, t)$  OR  $a \propto \exp(\sqrt{3 |\Lambda |} \, t)$?]

In summary we have:
\begin{itemize}
\item PBHs  produced before inflation have no observational
consequence, irrespective of the number of dimensions. 
\item $3+1$ PBHs cannot be produced during the inflation since the rate depends
on the $3+1$ Newtonian constant and is
small unless $\Lambda$ is Planckian.
\item  The $1+1$ dimensional phase is only ``effective'', the Universe
itself remaining $3+1$ dimensional.
\item If the $1+1$ dimensional phase persists through 
inflation, PBHs are produced prolifically since the rates
do not depend on the Planck mass, $G_{(2)}$ being dimensionless. However, their number density is exponentially diluted unless they are produced at the end of inflation.
\end{itemize}
We conclude that the decay of de Sitter universe has observational consequences only if the $(1+1)$-dimensional phase persisted until the end  of inflation. We do not necessarily regard this as a plausible situation, since one does not wish to risk sacrificing the attractions of the standard scenario. Nevertheless, we will confine attention to this possibility in what follows.     
%that survived during 

%THIS REPEATS  P13 BUT THE LATTER IS NOT SO EXPLICIT] 
In order to study the cosmological consequences of PBHs produced through this mechanism, we first recall the Hawking temperature $T(M)$  and production rate $\Gamma (M)$ for PBHs of mass $M$. For the ease of the presentation we set $G_{(2)}=1$.
For $M \gg \sqrt{|\Lambda |}$, which includes the sub-Planckian case for $\sqrt{|\Lambda |} \ll M_{\mathrm{P}}$, these are
\begin{equation}
T \approx M/(2 \pi), \quad \Gamma_\mathrm{L} \approx (\Lambda/M^2)^{1/2} \ll 1 \, .
\end{equation}
For $M = \sqrt{2 |\Lambda |}$, they are
\begin{equation}
T = \sqrt{|\Lambda |}/(\sqrt{2} \pi), \quad \Gamma_\mathrm{L} = 1 \, .
\end{equation}
For $ \sqrt{|\Lambda |} < M < \sqrt{2 |\Lambda |}$, they are
\begin{equation}
T \approx \sqrt{|\Lambda |}/(2 \pi), \quad \Gamma_\mathrm{L} \approx 1 \, .
\end{equation}
For $M = \sqrt{|\Lambda |} (1 + \epsilon)$ with $ \epsilon \ll 1$, they are
\begin{equation}
T \approx  \sqrt{ \epsilon |\Lambda |}/ (\sqrt{2} \pi) \ll \sqrt{|\Lambda |}, \quad \Gamma_\mathrm{L} \approx \epsilon^{-1/2} \gg 1 \, .
\end{equation}
For $M = \sqrt{|\Lambda |}$, corresponding to the Nariai black hole, they are
\begin{equation}
T = 0, \quad \Gamma_\mathrm{N} = (\mu_0/ \sqrt{|\Lambda |}) \, .
\label{nariaidm}
\end{equation} 
%\textcolor{red}{[The exponent $1/(4\pi)$ in these expressions should probably be $1/2$.]}  \textcolor{blue}{[Eq.(48) is probably $\Gamma = (\mu_o/ \sqrt{|\Lambda |})^{1/G_{(2)}}$ .]}
For $M < \sqrt{|\Lambda |}$, there is no black hole solution but a naked singularity. 
So $\Gamma(M)$ cuts off below the peak at $M = \sqrt{|\Lambda |}$ and has a power-law decline for $M \gg \sqrt{|\Lambda |}$.

The initial collapse fraction 
%$\beta(M) $
of PBHs of  mass $M$ is roughly
\begin{equation}
\beta (M) \sim \Gamma(M) \, ,
%/H(M) \sim \Gamma(M)/ \sqrt{|\Lambda|}
\end{equation}
with no $H(M)$ factor because $\Gamma$ is a probability rather than a rate.
 %s the Hubble rate when the PBHs form and 
%where $\Gamma (M)$ is given in Sec.~3.  
%[MORE PRECISE CALCULATION?] 
Since the scale factor in the de Sitter background scales as $\exp (t \sqrt{|\Lambda |})$, $H(M)$ is taken to be $\sqrt{|\Lambda | }$. After the PBHs have formed, the density of the de Sitter background is constant, whereas the PBH density  decreases as $a^{-1}$ in a $1+1$ model, so the fraction of the Universe's total (radiation) density in PBHs decreases as  
\begin{equation}
\rho_\mathrm{PBH}/ \rho_\mathrm{R} \propto a^{-1}  \propto \exp ( - t \sqrt{|\Lambda |}) \, ,
\end{equation}
where we have neglected evaporation. So the fraction decreases exponentially but not as fast as in the subsequent $3+1$ inflationary phase, where it decreases as $a^{-3}$.
\begin{figure}[h]
\begin{center}
\includegraphics[scale=0.65]{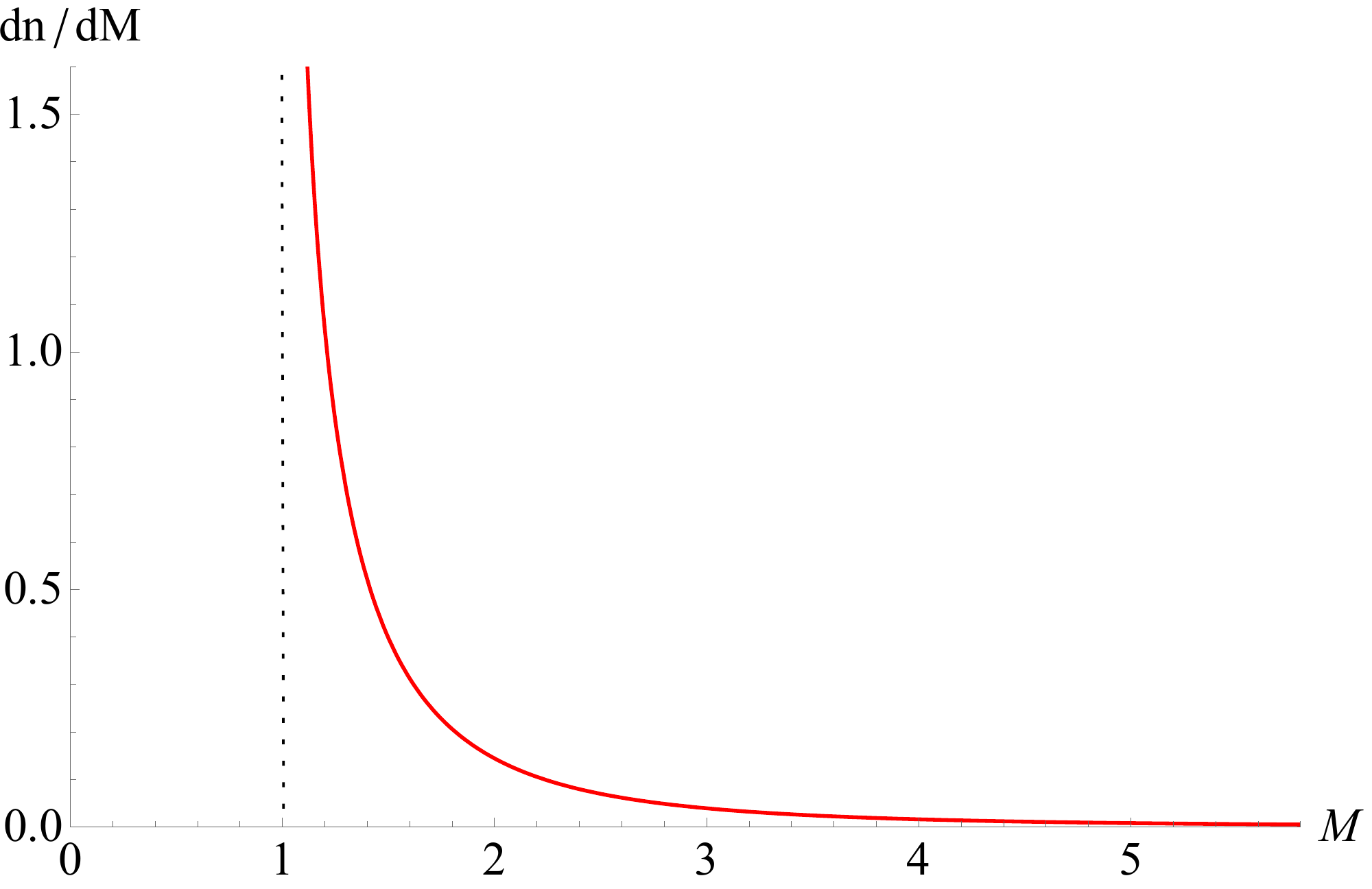}
\caption{Initial PBH mass function $\mathrm{d} n / \mathrm{d} M$, without the contribution from Nariai black holes. At the present epoch the mass function may have collapsed down to a delta-function at $M \sim M_{\rm CMB}$ due to evaporation.}
\label{fig:massfunction}
\end{center}
\end{figure}
The current PBH mass function should be
\begin{equation}
\frac{\mathrm{d} n}{\mathrm{d} M} \sim \frac{\beta(M)}{M^{2}} \sim \frac{\Gamma (M)}{ M^2}
%|\Lambda (M)|^{1/2}
\label{spectrum}
\end{equation} 
and this is shown in Fig. \ref{fig:massfunction}. 
The density of PBHs of mass $M$, denoted by $\rho(M)$,  is just $M^2$ times this and therefore comparable to $\Gamma (M)$. 
%BUT THIS OMITS THE CRUCIAL EXPONENTIAL DIILUTION - WE'RE  INTERESTED IN PBHS FORMING AT THE END OF INFLATION AND THIS MUST BE 1+1 INFLATION.] Like the production rate, it 
Both functions cut off below the peak at $M = \sqrt{|\Lambda |}$ and have a power-law decline for $M \gg \sqrt{|\Lambda |}$. The delta-function contribution from the Nariai black holes depends on $\mu_o$ and is therefore not included in Fig. \ref{fig:massfunction}.
There would be an exponential reduction given by eqn (50) if the black holes formed before the end of inflation and in this case the  PBHs would have no observational consequences at all. 
%[MASS FUNCTION  (58) PEAKS AT NARIAI MASS BUT DO THESE FORM AT END OF INFLATION? ]

We now consider two possible cosmological consequences of these black holes: (1) those with  $M \gg \sqrt{|\Lambda|}$ have a temperature $T \propto M$, so  their evaporation consequences are very different from those of $3+1$ black holes; (2) Nariai black holes with $M = \sqrt{|\Lambda|}$ have zero temperature and are therefore  stable, possibly contributing to the dark matter  (cf. the Planck mass relics of $3+1$ black holes if their evaporation stops at the Planck scale).  

\subsection{Evaporating sub-Planckian black holes}

%[THIS DISCUSSION IS PARTLY BASED ON REF~\cite{CMN15}] 
%THIS ANALYSIS OF WHAT HAPPENS AS A FUNCTION OF $M$ IS CORRECT BUT 
In this section, we discuss the evolution of a $1+1$ PBH of a specific mass $M$, leaving to the next section the issues of what value of $M$ might be expected and the effect of an extended mass function. We initially neglect the effect of the cosmic background radiation, which would suppress evaporation if it were hotter than the black hole, since this might not exist at early times, but we return to this point at the end. We also neglect the effect of accretion, which could also suppress evaporation if it were large, since this is expected to be small  \cite{CaH74}. 

From eqs.~(\ref{c3})  and (\ref{c5}), the black hole radius and temperature are 
%given by
\beq
R_\mathrm{S} \approx \frac{1}{2M} , \quad  T \approx \frac{M}{2 \pi} \, ,
\label{RT}
\eeq
%\PN{There is an inaccuracy here. From the horizon equation one gets $x_\mathrm{h}\approx 2M/|\Lambda|$ for $M\gg \sqrt{|\Lambda|}$. }
for $M \gg \sqrt{|\Lambda |}$, where $R_\mathrm{S}$ corresponds to what was previously called $x_\mathrm{h}$.
For $\sqrt{|\Lambda |} \ll M_\mathrm{P}$, the black holes can have less than the Planck mass and they would then resemble the sub-Planckian ones considered in Ref.~\cite{CMN15} in the $3+1$ context. In that work it was unclear whether black holes could form with sub-Planckian mass but here we have proposed a specific mechanism. Note that the condition $M \gg \sqrt{|\Lambda |}$ also allows $M > M_\mathrm{P}$, so (\ref{RT}) should  even apply in the super-Planckian regime. Indeed, for $\sqrt{|\Lambda |} \gg M_\mathrm{P}$, which may be unphysical, it could {\it only} apply in that regime.

In Ref.~\cite{CMN15} the luminosity of the black hole was written as
\beq
L 
\sim  R_\mathrm{S}^2 T^4 \sim \gamma  M^{2} \, ,
\label{lum}
\eeq
with $\gamma \equiv c^2/ \hbar$ and no dependence on $G_\mathrm{N}$.  This formula may seem  suspect in the present context since it assumes  the black hole is 3-dimensional, whereas  one expects the area to scale as $R_\mathrm{S}^{n-1}$ and the black-body emission to scale as $T^{1+n}$ with $n$ spatial dimensions. Curiously, however, this gives $L \propto M^2$ for $n=1$, so (\ref{lum}) still applies \cite{Mur12}. Indeed, since there is no $G_\mathrm{N}$ dependence in (\ref{lum}), the scaling $L \propto M^2$ is required on purely dimensional grounds.

\begin{figure}[h]
\begin{center}
\includegraphics[scale=0.35]{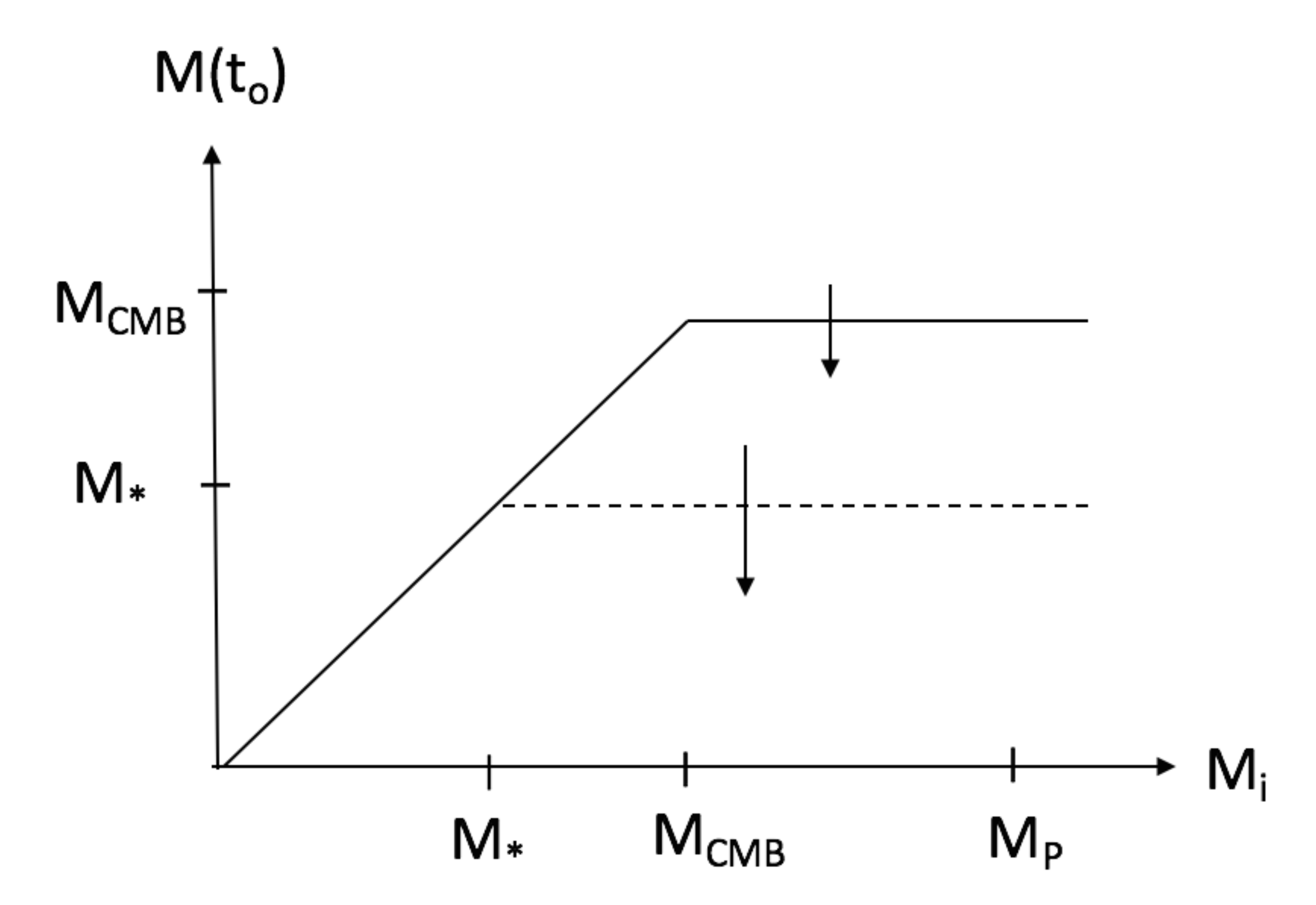}
\caption{Current black hole mass $M(t_0)$ as a function of initial mass $M_i$, showing mass $M_*$ for which evaporation timescale equals age of Universe and mass $M_{\rm CMB}$ for which  black hole has current CMB temperature. Both mass-scales decrease with time but the latter does so more slowly.}
\label{fig8}
\end{center}
\end{figure}

If the black hole forms at time $t_\mathrm{i}$ with mass $M_\mathrm{i}$, then (\ref{lum}) implies its mass subsequently  evolves according to 
\beq
t - t_\mathrm{i} 
\sim  \frac{1}{\gamma}  \left( \frac{1}{M} - \frac {1}{M_\mathrm{i}} \right) \, .
\label{evap}
\eeq
Although it decreases on a `Compton' timescale, 
\beq
\tau  \sim 1/(\gamma M)  \sim \hbar/(Mc^2) \, ,
\eeq
the black hole never evaporates entirely because Eq.~(\ref{evap}) shows that it takes an infinite time for $M$ to reach zero. 
%Nevertheless, 
However, we note that there is a value of $M$ for which $\tau$ is comparable to the age of the Universe ($t_0 \sim 10^{17}$s) and this is
\beq
M_{*} \sim  1/(\gamma t_0) \sim \hbar / (c^2 t_0)  \sim 10^{-65}~\rm g \, .
\eeq
From (\ref{evap}), the mass of the black hole at the present epoch is then
\beq
M = \frac{M_\mathrm{i}}{1+ \gamma M_\mathrm{i}(t_o - t_\mathrm{i})} \approx\frac{M_\mathrm{i}}{1+ M_\mathrm{i}/M_{*}}  \, .
\eeq
Hence $M \approx M_\mathrm{i}$ for $M_\mathrm{i} \ll M_{*}$ (i.e. the mass is unchanged) but $M \approx M_{*}$ for $ M_i \gg M_{*}$. This is indicated by the dotted line in Fig.~\ref{fig8}.

This mass-scale $M_*$ is associated with  a radius of $10^{27}$cm (the current cosmological horizon size) and a temperature of $10^{-28}$K (the Hawking temperature for a black hole with the mass of the Universe). 
It  might  seem implausibly small but this mass-scale arises naturally in some estimates for the photon or graviton mass (e.g. 
in the work of Mureika and Mann~\cite{rbmjm}). It might also  have observational consequences associated with gravitational effects on the scale of clusters and the Dvali-Gabadadze-Porrati (DPG) effect.
%It effectively specifies the lower integration bound in (\ref{entropy}), {\it i.e.} $M_{**} = M_0$.
% \textbf{[PN: question. If feel that the last condition should be for $M_* \gg M_\mathrm{i} \gg M_{**}$. As it is $M$ is simultaneously $M \gg M_{**}$ and $M \approx M_{**}$]}

The above analysis neglects the effect of the cosmic microwave background (CMB). This may be appropriate until the end of inflation, since there may be no background radiation then. However, the assumption  fails after reheating and evaporation will be suppressed whenever the black hole temperature is less than the CMB temperature ($T_{\rm CMB}$). This means that the PBH mass may never actually reach the tiny value $M_{*}$.
 %because of thae effect of the cosmic microwave background (CMB). This is because the black hole temperature is less than the CMB temperature ($T_{\rm CMB}$), suppressing evaporation altogether, 
Indeed, the CMB should prevent PBH evaporation below an epoch-dependent mass
\beq
M_{\rm CMB} = 10^{-36}(T_{\rm CMB}/3K) \,  \mathrm{g} \, .
\eeq
%FIG 8 IS CORRECT BUT NEEDS MORE EXPLANATION HERE.] 
Since accretion is expected to be unimportant \cite{CaH74}, the PBH mass should effectively freeze at this value, \textit{i.e.}, its value can be approximated as $M_{\rm i}$ for $M_{\rm i}<M_{\rm CMB}$ and $M_{\rm CMB}$ for $M_{\rm i} > M_{\rm CMB}$. This is indicated by the solid line  in Fig.~\ref{fig8}. 
Note that $M_* $ and $M_{\rm CMB}$ both decrease with time but as $t^{-1}$ and  $t^{-2/3}$. respectively,  so $M_*$ falls faster.
%[ONE REALLY NEEDS TO INTEGRATE OVER THE MASS FUNCTION (\ref{spectrum}) AND NARIAI ONES MAY DOMINATE.]
 %[MODIFIED]. 
% and DGP \textbf{[we should define this acronym]} effects ~\cite{nieto}. 

\subsection{Lower dimensional PBHs as dark matter}

The above scenario leads to stable relics which might in principle provide the  dark matter.
%~\cite{poly_1,poly_2,poly_3,poly_4,poly_6,poly_7,jrmbh}. 
The relic PBH mass decreases with time but the current value is around $10^{-4}$eV, which is close to the mass-scale associated with the dark energy. This coincidence reflects the fact that the dark energy density ($\rho_{\rm DE} \sim \Lambda_\mathrm{c}$) and CMB density ($\rho_{\rm CMB} \sim T_{\rm CMB}^4 \sim M_{\rm CMB}^4$) are not so different at the present epoch, corresponding to just a factor of $10$ in $M_{\rm CMB}$.
 
We have  seen that the PBHs formed in this scenario can only have an appreciable density if they form at the end of inflation. In this case, eqn.~(\ref{spectrum}) should still apply but with the value of the cosmological constant at the end of inflation ($\Lambda_i$). The dominant contribution to the density would then come from PBHs with initial mass $\sqrt{\Lambda_i}$. Providing this exceeds the mass $M_{\rm CMB}$, the analysis of Sec.~5.1 suggests that all these PBHs will have shrunk to $M_{\rm CMB}$, with the mass function shown in Fig.~\ref{fig:massfunction} turning into a delta-function at around that mass. 
Otherwise they will still have the mass $\sqrt{\Lambda_i}$.

We must also consider the effects of the Nariai black holes, these necessarily having a current  mass $\sqrt{\Lambda_i}$ since they do not evaporate at all. From eqn.~(\ref{spectrum}), their density at formation is 
\beq
\rho_ {\mathrm{N}} (\Lambda_i) \sim M^2 \,  \frac{ \mathrm{d} n_{\mathrm{N}}}{\mathrm{d} M} \sim \Gamma_{\mathrm{N}} (\Lambda_i) \sim \frac{\mu_0}{\sqrt{\Lambda_i}} \, .
%)}{|\Lambda|^{1/2}} \, .
\eeq
%with no reduction due to evaporation. 
(Strictly, one should integrate over $\Lambda$ but the dominant effect clearly comes from $\Lambda_i$ due to the exponential dilution prior to reheating.) The current PBH mass function should therefore comprise  two delta-functions, one at  $M \sim \sqrt{\Lambda_i}$ and the other at $M \sim M_{\rm CMB}$. This raises the question of 
%whether Nariai black holes 
which component could most plausibly provide the dark matter. The Nariai contribution  depends on the value of $\mu_0$ in (\ref{nariaidm}), this being essentially a free parameter. 
%The current density is smaller by a factor $(R_o/R_i)^3$, so
Of course,  it requires very fine-tuning of $\Lambda_i$, $t_i$ and $\mu_0$ to explain the dark matter but such fine-tuning is a feature of all PBH scenarios. Having  the dark matter in ($1+1$)-dimensional objects with zero temperature might seem  rather radical but one does expect the PBH mass function to peak at this mass.
%[EXPAND. CAN WE IDENTIFY $\Lambda_i$ WITH $\Lambda_C$? ]

\section{Summary and future work}
\label{conclusion}

In this paper we have studied the spontaneous production of  PBHs if the Universe was effectively  $(1+1)$-dimensional before it became $(3+1)$-dimensional. We have investigated this quantum nucleation process semi-classically by using the instanton method
and constructing  instantons which  represent an early $(1+1)$-dimensional universe.  
%After fixing the form  of the dilatonic function for each universe, we estimated  
Our estimate of the PBH creation rate suggests that they could be very abundant 
%in the 1+1 phase 
and have an initial mass 
%Nariai  value 
$\sim \sqrt{|\Lambda|} $ and size $\sim 1/ \sqrt{|\Lambda|}$, where $\Lambda$ is the value of the cosmological constant in the (inflating) $1+1$ background.  The ones which form at the end of the inflationary phase will dominate the current density. Providing $\sqrt{|\Lambda _i |}$ exceeds the mass of the 
%cosmic microwave background 
CMB photons, the PBHs will decay down to that mass,  leaving stable relics which could provide the dark matter. 
%have positive heat capacity and that makes them are thermodynamically stable. So, their lifetime is infinite and this suggests that their production might have significantly influenced  the  structure formation and density distributions in the early universe. 
Alternatively, the dark matter could be associated with Nariai black holes which retain the mass  $\sqrt{|\Lambda _i |}$ because they never evaporate at all. If one identifies $|\Lambda _i |$ with the current 4D cosmological constant, this naturally explains why the dark energy and dark matter have comparable densities.  
For this model to be viable, the $1+1$ phase must persist until the end of inflation and we suggest a scenario in which  this happens plausibly in an accompanying paper.

\subsection*{Acknowledgments}

The work of P.N. has been supported by the project ``Evaporation of the microscopic black holes'' of the German Research 
Foundation (DFG) under the grant NI 1282/2-2 and by the Helmholtz International Center for
FAIR within the framework of the LOEWE program (Landesoffensive zur Entwicklung
Wissenschaftlich-Ökonomischer Exzellenz) launched by the State of Hesse.  A.G.T, P.N. and J.M. are grateful to the Max Planck Institute for Radioastronomy, Bonn, for hospitality during the early stages of this work.

% Activate the appendix
% from now on sections are numerated with capital letters
\appendix

\section{Robertson-Walker metric in two dimensions}\label{Sec:appendixA}

To map the two-dimensional  de Sitter solution, 
%\eqref{b5}, namely
\begin{equation} \label{e1}
\mathrm{d} s^2 = - (1- |\Lambda| x^2) \mathrm{d} t^2 + \frac{\mathrm{d} x^2}{(1- |\Lambda| x^2)} \,,
\end{equation}
into the two-dimensional Robertson-Walker metric,
\begin{equation} \label{e2}
\mathrm{d} s^2 = - \mathrm{d} t^2 + a^2(t) \mathrm{d} x^2 \, ,
\end{equation}
one performs the following coordinate transformation:
%$(t,x) \rightarrow (T,r)$:
\begin{eqnarray} 
t &=& T - \frac{1}{2\sqrt{|\Lambda|}} \ln \left( -\frac{1}{|\Lambda|} + r^2 e^{2\sqrt{|\Lambda|} T} \right) \,, \\ \label{e4}
x &=& r e^{\sqrt{|\Lambda|} \, T} \,.
\end{eqnarray}
After differentiation, one gets
\begin{eqnarray} \label{e5}
\mathrm{d} t &=& \frac{\sqrt{|\Lambda|}}{1-|\Lambda| r^2 e^{2\sqrt{|\Lambda|} \, T}} \left( \frac{\mathrm{d} T}{\sqrt{|\Lambda|}}  + r e^{2\sqrt{|\Lambda|} T} \mathrm{d}r \right) \,, \\ \label{e6}
\mathrm{d} x &=& e^{\sqrt{|\Lambda|} T} \mathrm{d} r + r \sqrt{|\Lambda|} e^{\sqrt{|\Lambda|} \, T} \mathrm{d}T \,.
\end{eqnarray}
Putting \eqref{e4}, \eqref{e5} and  \eqref{e6} into \eqref{e1} gives
\begin{equation} \label{e7}
\mathrm{d} s^2 = - \mathrm{d} T^2 + e^{2\sqrt{|\Lambda|} \, T} \mathrm{d} r^2 \,.
\end{equation}
%By renaming 
With the relabelling $T \rightarrow t$ and $r \rightarrow x\,$, 
%\eqref{e7} 
this becomes 
%of the same form of \eqref{e2}:
\begin{equation} \label{e8}
\mathrm{d} s^2 = - \mathrm{d} t^2 + e^{2\sqrt{|\Lambda|} \, t} \mathrm{d} x^2 \,,
\end{equation}
allowing us to identify the scale factor in (\ref{e2})  as
\begin{equation} \label{e9}
a(t) = e^{\sqrt{|\Lambda|}\,  t} \,.
\end{equation}
%[SIMILAR ANALYSIS FOR SCHWARZSCHILD-DE SITTER SOLUTION?]

We can also map the Nariai metric
\if
, given by eq.(41) which is the same as eq.(19) coming from deSitter space, i.e,
\begin{equation} \label{e1}
\mathrm{d} s^2 = \frac{1}{|\Lambda|} \left( \mathrm{d} \chi^2 + \sin^2 \chi \ \mathrm{d} \xi^2 \right)  \,,
\end{equation}
into the cosmological $1+1$ FRW metric, which is given by
\begin{equation} \label{e2}
\mathrm{d} s^2 = - \mathrm{d} t^2 + \alpha(t)^2 \mathrm{d} x^2 \,,
\end{equation}
\fi
into the  $1+1$ FRW metric
by introducing the coordinates
%making the following coordinate transformation ( $(\xi,\chi) \rightarrow (T,r)$):
\begin{eqnarray} %\label{e3}
\xi &=& i \sqrt{|\Lambda|} \left(  T - \frac{1}{2\sqrt{|\Lambda|}} \ln \left( -\frac{1}{|\Lambda|} + r^2 e^{2\sqrt{|\Lambda|} T} \right) \right)  \,, \\ %\label{e4}
\chi &=& \arccos \left( \sqrt{|\Lambda|} r e^{\sqrt{|\Lambda|}T} \right) \,.
\end{eqnarray}
Differentiating then gives
\begin{eqnarray} %\label{e5}
\mathrm{d} \xi &=& i \frac{\sqrt{|\Lambda|} \ \mathrm{d} T + |\Lambda| r e^{2 \sqrt{|\Lambda|} T } \mathrm{d} r}{1- e^{2 \sqrt{|\Lambda|} T }|\Lambda| r^2 } \,, \\ %\label{e6}
\mathrm{d} \chi &=& - \frac{\sqrt{|\Lambda|} e^{\sqrt{|\Lambda|}T}}{\sqrt{1- e^{2 \sqrt{|\Lambda|} T }|\Lambda| r^2}} \left( \mathrm{d} r + r \sqrt{|\Lambda| } \mathrm{d} T  \right)  \,.
\end{eqnarray}
Substituting these relations
%Plugging \eqref{e4}, \eqref{e5} and  \eqref{e6}
 into \eqref{b9} again yields metric \eqref{e8}. 
Note that the cosmological  background can itself be regarded as a black hole in the Nariai solution.
\if
%\begin{equation} \label{e7}
%\mathrm{d} s^2 = - \mathrm{d} T^2 + e^{2\sqrt{|\Lambda|} T} \mathrm{d} r^2 \,.
%\end{equation}
%The relabelling $T \rightarrow t$ and $r \rightarrow x\,$, 
eq.\eqref{e7} becomes of the same form with eq.\eqref{e2}:
\begin{equation} \label{e8}
\mathrm{d} s^2 = - \mathrm{d} t^2 + e^{2\sqrt{|\Lambda|} t} \mathrm{d} x^2 \,,
\end{equation}
again allows us to identify the scale factor 
%from eq.\eqref{e2} for
in  the $1+1$ Nariai black hole with  Eq.~(\ref{e9}). 
%\begin{equation} \label{e9}
%\alpha(t) = e^{\sqrt{|\Lambda|} t} \,,
%\end{equation}
%just like the scale factor of $1+1$-D deSitter space.
%[THIS IS SURELY OBVIOUS SINCE IT'S THE SAME METRIC AS ABOVE BUT WHY DOES IT NOT CONTAIN A BLACK HOLE?] 
\fi

\end{document}